\documentclass{svproc}
\usepackage{url}

\usepackage{hyperref}
\newcommand\modelabbrv{HM-LDM}
\newcommand\modelname{Hybrid-Membership Latent Distance Model}
\usepackage{cite}
\usepackage{amsmath,amssymb,amsfonts}
\usepackage{algorithmic}
\usepackage[pdftex]{graphicx}
\usepackage{bm}
\usepackage{booktabs}  
\usepackage{subfig}
\usepackage{xcolor}
    
\begin{document}
\mainmatter              
\title{\modelabbrv: A \modelname}

\titlerunning{\modelabbrv}  
%
\author{Nikolaos Nakis \and Abdulkadir Çelikkanat \and Morten Mørup}
%

%
%
%
\tocauthor{Nikolaos Nakis, Abdulkadir Çelikkanat, Morten Mørup}
\institute{Section for Cognitive Systems,\\Technical University of Denmark, Kongens Lyngby 2800, Denmark\\
\email{nnak@dtu.dk, abce@dtu.dk, mmor@dtu.dk}}

\maketitle              

\begin{abstract}
A central aim of modeling complex networks is to accurately embed networks in order to detect structures and predict link and node properties. The Latent Space Model (LSM) has become a prominent framework for embedding networks and includes the Latent Distance Model (LDM) and Eigenmodel (LEM) as the most widely used LSM specifications. For latent community detection, the embedding space in LDMs has been endowed with a clustering model whereas LEMs have been
constrained to part-based non-negative matrix factorization (NMF) inspired
representations promoting community discovery. We presently reconcile LSMs with latent community detection by constraining the LDM representation to the $D$-simplex forming the Hybrid-Membership Latent Distance Model (\textsc{\modelabbrv}). We show that for sufficiently large simplex volumes this can be achieved without loss of expressive power whereas by extending the model to squared Euclidean distances, we recover the LEM formulation with constraints promoting part-based representations akin to NMF. Importantly, by systematically reducing the volume of the simplex, the model becomes unique and ultimately leads to hard assignments of nodes to simplex corners. We demonstrate experimentally how the proposed \textsc{\modelabbrv} admits accurate node representations in regimes ensuring identifiability and valid community extraction. Importantly, \textsc{\modelabbrv} naturally reconciles soft and hard community detection with network embeddings exploring a simple continuous optimization procedure on a volume constrained simplex that admits the systematic investigation of trade-offs between hard and mixed membership community detection.
\end{abstract}
\setlength{\parskip}{0pt}

\section{Introduction}\label{sec:introduction}
Networks naturally arise in the vast majority of scientific domains from physics to biology in order to model interactions among diverse entities with numerous instances such as collaboration, protein-protein, and brain connectivity networks \cite{newman}. Hence, graph analysis tools have become crucial to extract and analyze the underlying meaningful information from networks. In this direction, Graph Representation Learning (GRL) \cite{GRL-survey-ieeebigdata20} approaches have become a dominant way to carry out various downstream tasks such as node classification, link prediction, and community detection thanks to their superior performance compared to the classical techniques. GRL models mainly aim to map similar nodes in the network to close latent positions in a low dimension space by automatically learning corresponding node features \cite{survey_hamilton_leskovec}. 

The initial GRL works aimed to learn representations or features by simulating random walks over networks, taking inspiration from the Natural Language Processing field \cite{deepwalk-perozzi14, node2vec-kdd16, expon_fam_emb, netmf-wsdm18, line}. They mainly extract embeddings by optimizing the co-occurrence probability of node pairs within a certain distance through random walks. In recent years, we have witnessed a tremendous increase in the number of Graph Neural Networks (GNN) \cite{survey_hamilton_leskovec} methods with their usage in supervised tasks. They primarily rely on iterative message-passing operations of node attributes and hidden features around the surroundings of nodes for a given task. The matrix decomposition-based models \cite{netmf-wsdm18, netsmf-www2019} are also a notable class of the GRL methods. They learn node embeddings by decomposing a designed target matrix based on first and higher-order proximity. However, few GRL methods rely on Non-negative Matrix Factorization (NMF), although it is a popular technique for unsupervised signal decomposition and approximation of multivariate non-negative data. NMF techniques have gathered lots of attention since they allow for structure retrieval through the latent factors of the imposed decomposition providing easy interpretable part-based representations\cite{lee99}.

Applications of NMF include network analysis allowing for efficient, unsupervised, and overlapping community detection, as well as GRL \cite{nmf1,nmf2,nmf3,nmf4}. Within the NMF formulation, various works have sought to define mixed-membership frameworks for analysis and community detection purposes. A Mixed-Membership Stochastic Block Model (MM-SBM) \cite{JMLR:v9:airoldi08a} has been linked to the symmetric-NMF decomposition with uniqueness guarantees \cite{nmf4}. 
Standard least-squares NMF optimization was exchanged to a Poisson likelihood optimization for obtaining the propensity of nodes belonging to different communities \cite{nmf1}. In addition, a GRL approach for overlapping communities was presented in \cite{nmf2} where NMF was utilized to discover Poisson distributed mixed-memberships. These works, design mixed-memberships vectors for part-based representations \cite{lee99} projected in an NMF constructed space where node similarity, as well as, position and metric properties, can be abstract.

The Latent Space Models (\textsc{LSM}s) are also one of the most powerful ways to learn latent representations \cite{nakis2022hierarchical}. These methods employ generalized linear models for constructing latent node embeddings which express important network characteristics. More specifically, the \textsc{LDM} \cite{exp1} employs the Euclidean norm for positioning similar nodes closer in the latent space, which comes as a direct consequence of the triangular inequality, naturally representing transitivity ({\it"a friend of a friend is a friend"}) and homophily ({\it a tendency where similar nodes are more likely to connect to each other than dissimilar ones}) properties. The \textsc{LDM} can be generalized through the Eigenmodel \cite{hoff2007modeling} that can account for stochastic equivalence ({\it"groups of nodes defined by shared intra- and inter-group relationships"}) akin to the SBM \cite{JMLR:v9:airoldi08a} and the mixed membership SBM \cite{JMLR:v9:airoldi08a}.
Furthermore, \textsc{LDM}s have been endowed with a clustering model imposing a Gaussian Mixture Model as prior forming the latent position clustering model \cite{handcock2007model,ryan2017bayesian}.

In this study, we propose a novel unsupervised representation learning method over graphs, namely, the \modelname \ (\textsc{\modelabbrv}), by bringing together the strengths of LDM and NMF. 
Specifically, the \textsc{\modelabbrv} offers a reconciliation between part-based representations of networks and low-dimensional latent spaces satisfying similarity properties such as homophily and transitivity. The choice of these similarity properties is of high significance and one of the key characteristics behind GRL since they allow for easily interpretable discovery of network structure. Additionally, our proposed method permits us to capture the latent community structure of the networks using a simple continuous optimization procedure over the log-likelihood of the network. Notably, unlike most existing approaches imposing hard community memberships constraints, the assignment of community memberships in our proposed 
hybrid model can be controlled and altered through the simplex volume formed by the latent node representations. We extensively evaluate the performance of the proposed method in the ability to perform link prediction, as well as, community discovery over various networks of different types. We demonstrate that our model outperforms recent methods.


\noindent\textbf{Source code:} \href{https://github.com/Nicknakis/Hybrib-Membership-Latent-Distance-Model}{\textit{Hybrid-Membership Latent Distance
Model}}.

\section{Problem statement and proposed method}\label{sec:method}
Let $\mathcal{G}=(\mathcal{V},\mathcal{E})$ be an undirected graph where $\mathcal{V}$ shows the vertex set and $\mathcal{E} \subseteq \mathcal{V}\times \mathcal{V}$ the edge set. We use $\mathbf{Y}_{N \times N}=\left(y_{i,j}\right)\in \{0,1\}^{N\times N}$ to denote the adjacency matrix of the graph where $y_{i,j}=1$ if the pair $(i,j) \in \mathcal{E}$ otherwise it is equal to $0$ for all $ 1\leq i< j\leq N := |\mathcal{V}|$. Our main goal is to learn a representation, $\mathbf{w}_i \in \mathbb{R}^{D}$, for each node $i \in \mathcal{V}$ in a lower dimensional space ($D \ll N$) such that similar nodes in the network should have close embeddings. More specifically, we concentrate on mapping the nodes into the unit $D$-simplex set, $\Delta^{D} \subset \mathbb{R}_{+}^{D+1}$.
Therefore, the extracted node embeddings can convey information about latent community memberships. Many GRL approaches also do not provide identifiable or unique solution guarantees, so their interpretation highly depends on the initialization of the hyper-parameters. In this study, we will also address the identifiability problem and seek identifiable solutions which can only be achieved up to a permutation invariance, as reported in Def. \ref{def:identifiabilty}. 

\begin{definition}[\textbf{Identifiabilty}]\label{def:identifiabilty}
An embedding matrix $\mathbf{W}$ whose rows indicating the corresponding node representations is called an \textit{identifiable solution up to a permutation} if it holds $\widetilde{\mathbf{W}}=\mathbf{W}\mathbf{P}$ for a permutation $\mathbf{P}$ and a solution $\widetilde{\mathbf{W}} \not= \mathbf{W}$.
\end{definition}

We define a Poisson distribution over the adjacency matrix $\mathbf{Y}$ of the network $\mathcal{G}=(\mathcal{V}, \mathcal{E})$ to be conditionally independent given the unobserved latent positions, and write the log-likelihood function as follows:
\begin{equation}
    \label{eq:prob_adj}
    \log P(\mathbf{Y}|\bm{\Lambda})=\!\!\sum_{\substack{i<j \\ y_{ij}=1}}\!\log(\lambda_{ij})\;-\;\sum_{\substack{i< j }}\Big(\lambda_{ij}+\log(y_{ij}!)\Big) \:.
\end{equation}
where $\bm{\Lambda}=(\lambda_{ij})$ is the Poisson rate matrix which has absorbed the dependency over the model parameters. We here adopted a Poisson regression model similar to the work in \cite{doi:10.1198/016214504000001015}. In this study, we make use of a Poisson likelihood for modelling binary networks, as validated in \cite{nmf2}.

We propose the \modelname \ (\textsc{\modelabbrv}) with a log-rate based on the $\ell^2$-norm as:
  \begin{equation}
     \label{eq:nmf_rate}
     \log \lambda_{ij}=\Big(\gamma_i+\gamma_j-\delta^p\cdot||\mathbf{w}_i -\mathbf{w}_j||_2^p\Big),
 \end{equation}
 where $\mathbf{w_i} \in [0,1]^{D+1}$ and $\sum_{d=1}^{D+1} w_{id}=1$, $\delta \in \mathbb{R}_+$ and $\gamma_i \in \mathbb{R}$ denotes the node-specific random-effects \cite{doi:10.1198/016214504000001015,KRIVITSKY2009204} describing essentially the tendency of nodes to sending and receiving connections, accounting for degree heterogeneity. In addition, the norm degree $p \in \{1,2\}$ controls the power of the $\ell^2$-norm and combined with the latent embeddings sum-to-one condition constrains the latent space to the $D-$simplex with size equal to $\delta$. 
 A remarkable property of Eq. \eqref{eq:nmf_rate}, for $p=2$, is that it resembles a positive Eigenmodel with random effects: $ \tilde{\gamma}_i+\tilde{\gamma}_j+(\mathbf{\tilde{w}}_i\bm{\Lambda}\mathbf{\tilde{w}}_j^{\top})$ where $\bm{\Lambda}$ is a diagonal matrix having non-negative elements, i.e. $\tilde{\gamma}_i=\gamma_i-\delta^2\cdot||\mathbf{w}_i||^2_2$, $\tilde{\gamma}_j=\gamma_j-\delta^2\cdot||\mathbf{w}_j||^2_2$ and $\tilde{\mathbf{w}}_i\bm{\Lambda}\tilde{\mathbf{w}}_j^\top=2\delta^2\cdot \mathbf{w}_i\mathbf{w}_j^\top$ thus the squared Euclidean distance reconciles the conventional LDM and non-negativity constrained Eigenmodel. The squared Euclidean distance is not fully a metric but it still expresses homophily, leading to an interpretable latent space. Even though the triangle inequality is not exactly satisfied, it preserves the ordering of pairwise Euclidean distances, and it is highly preferred in applications since it is a strictly convex smooth function. By the well-known cosine formula, we have
\begin{align}
||\mathbf{w}_i-\mathbf{w}_j||_2^2 &= ||\mathbf{w}_i-\mathbf{w}_k||_2^2+|| \mathbf{w}_k-\mathbf{w}_j ||_2^2+2||\mathbf{w}_i-\mathbf{w}_k||_2||\mathbf{w}_k-\mathbf{w}_j)||_2\cos(\theta),\nonumber
\end{align}
\noindent where $\theta \in (-\pi/2, \pi/2)$ is the angle between the vectors $\mathbf{w}_i-\mathbf{w}_k$ and $\mathbf{w}_k-\mathbf{w}_j$. Notice that the third term also converges to $0$ for similar nodes since we will have close representations. For the case where $\theta \in [-\pi, -\pi / 2]\cup[\pi/ 2, \pi]$, it holds the triangle inequality: $||\mathbf{w}_i - \mathbf{w}_j ||_2^2 \leq || \mathbf{w}_i - \mathbf{w}_k ||_2^2 + || \mathbf{w}_k - \mathbf{w}_j ||_2^2$. 

\begin{figure}[!t]
  \centering
    \subfloat[Translation invariances.]{{
  \includegraphics[width=0.21\columnwidth]{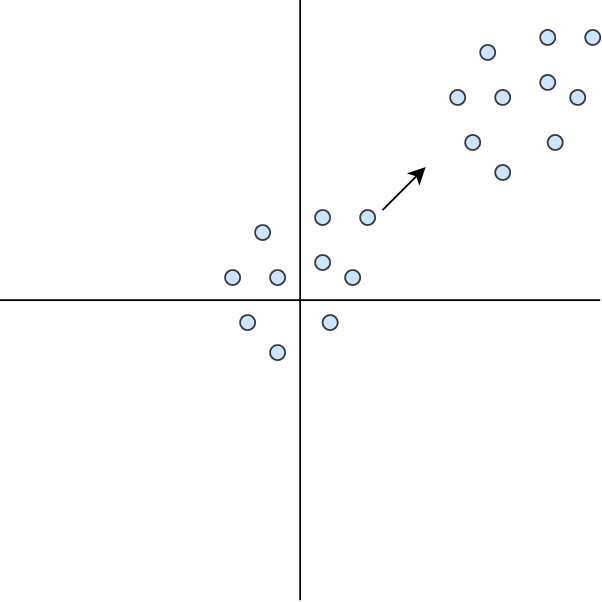} }}%
  \hfill
   \subfloat[Rotation invariances.]{{
  \includegraphics[width=0.25\columnwidth]{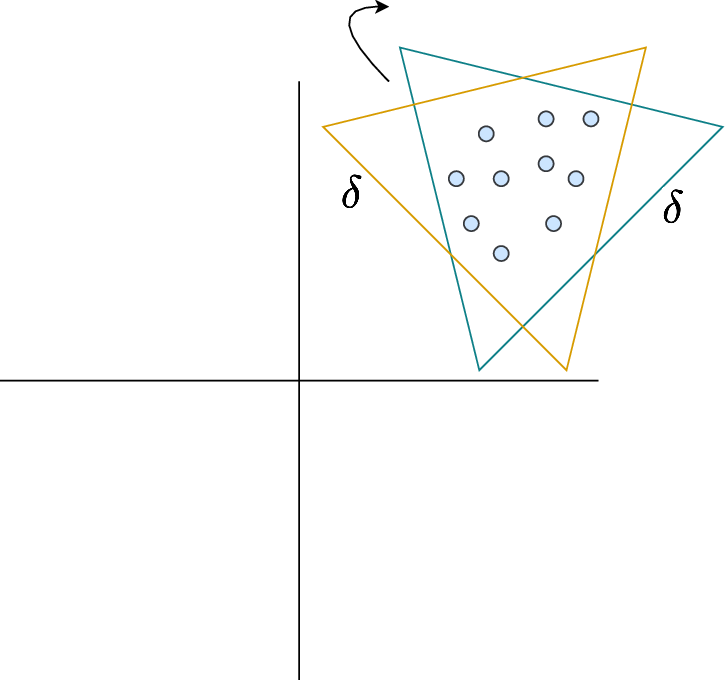} }}%
  \hfill
  \subfloat[Decreased simplex volume ensuring identifiability.]{{\includegraphics[width=0.34 \columnwidth]{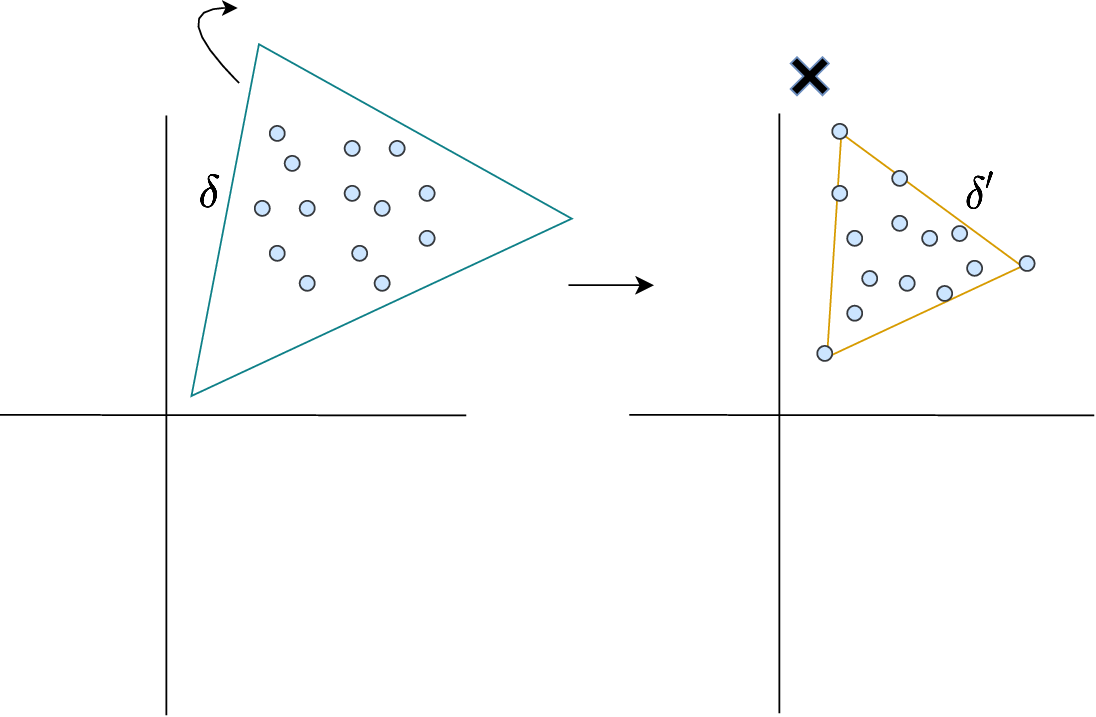} }}%
  \caption{A $2$-dimensional latent space with the $2$-simplex given as the green and yellow triangles, the blue points denote embedding positions of the $\textsc{LDM}$ and $\delta$ is the simplex size.}\label{fig:invariances}
\end{figure}

The embedding vectors, $\{\mathbf{w}_i\}_{i=1}^{N}$ in Eq. \eqref{eq:nmf_rate}, are constrained to non-negative values and to sum to one. Thereby, they reside on a simplex showing the participation of node $i\in\mathcal{V}$ over $D+1$ latent communities. Any \textsc{LDM} can be translated to the non-negative orthant without any loss in performance or in expressive capability. Non-negative embeddings do not affect the distance metric, as it is invariant to translation, as shown by Figure \ref{fig:invariances} (a). In addition, the $D$-dimensional non-negative orthant can be reconstructed by a large enough $D$-simplex. Based on these arguments, it is trivial to show that for large values of the $\delta$ parameter in Eq. \eqref{eq:nmf_rate}, despite the sum-to-one constraint on the embeddings $\mathbf{W}$, we obtain an unconstrained \textsc{LDM}, as distances are unbounded when $\delta\rightarrow+\infty$. In this case, the memberships defined by $\mathbf{W}$ are not uniquely identifiable due to the distance invariance of rotation, as seen in Figure \ref{fig:invariances} (b).
 However, by shrinking the volume of the simplex (equivalent to decreasing $\delta$), eventually the $D$-dimensional space of \textsc{LDM} will no longer be enclosed inside the $D$-simplex, forcing nodes to start populating the corners of this smaller simplex. We call a node \textit{champion} if its latent representation is a standard binary unit vector.

\begin{definition}[\textbf{Community champion}]\label{def:champion}
A node for a latent community is called \textit{champion} if it belongs to the community (simplex corner) while forming a binary unit vector.
\end{definition}
 
The champion nodes are of great significance for identifiability because if every corner of the simplex is populated by at least one node (champion), then the solution of the model is identifiable (up to a permutation matrix) (Def. \ref{def:identifiabilty}) as any random rotation does not leave the solution invariant anymore, as shown by Figure \ref{fig:invariances} (c). We observe then, that the scalar, $\delta$, controls the type of memberships of the model and its expressive capabilities. Large enough values lead to the basic \textsc{LDM} but inherits its rotational invariance. Small values of $\delta$ lead to identifiable solutions and ultimately hard cluster assignments. Thereby, for very small values of $\delta$, nodes are solely assigned to the simplex corners. Lastly, we can also find regimes of values for $\delta$ that offer identifiable solutions but also performance similar to \textsc{LDM}, defining a silver lining.

A different take on the identifiability of the model for $p=2$, can also be given under the Non-negative Matrix Factorization (NMF) theory. This is easily shown by a re-parameterization of Eq. \eqref{eq:nmf_rate} by $\tilde{\gamma}_i+\tilde{\gamma}_j+2\delta^2\cdot(\mathbf{w}_i\mathbf{w}_j^{\top})$ as described in Eq. \eqref{eq:nmf_rate}. In this formulation, the product $\mathbf{W}\mathbf{W}^{\top}$ defines a symmetric NMF problem which is an identifiable and unique factorization (up to permutation invariance) when $\mathbf{W}$ is full-rank and at least one node resides solely in each simplex corner, ensuring separability \cite{nmf4,nmf5}. Under this NMF formulation, the product $\mathbf{w}_i\mathbf{w}_j^{\top} \in [0,1]$ achieves its upper bound only if both nodes $i$ and $j$ reside in the same corner of the simplex. The parameter, $\delta$, acts as a simple multiplicative factor in the first term of the objective function of \textsc{\modelabbrv}, given in Eq. \eqref{eq:prob_adj}, while in the second term acts as a power of the exponential function. For small values of $\delta$, the model is biased towards hard latent community assignments of nodes since similar nodes achieve high rates only when they belong to the same latent community (simplex corner). On the other hand, nodes heading towards the simplex corners for large values of $\delta$ lead to an exponential change in the second term of the log-likelihood function given in Eq. \eqref{eq:prob_adj}. Thus, a possible hard allocation of dissimilar nodes to the same community penalizes the likelihood severely. For this reason, high order of $\delta$ benefits mixed-membership allocations.
  
\section{Experimental evaluation}\label{sec:experimental_evaluations}
We proceed by evaluating the efficiency and performance of the proposed method. 
In our set-up, we make use of networks with unknown community structures, as well as, with ground-truth communities. We employ the former networks to validate the ability of our framework to discover identifiable latent structures and predict missing links. The latter networks are used to verify that the $\textsc{\modelabbrv}$ discovers communities successfully. We consider multiple social and scientific collaboration networks as shown by Table \ref{tab:network_statistics}. We treat all networks as unweighted and undirected.

For the training of \textsc{HM-LDM} we optimize the log-likelihood function of Eq. \eqref{eq:prob_adj} via the Adam optimizer \cite{kingma2017adam} with learning rate $lr \in [0.01,0.1]$. The node-specific random effects vector $\bm{\gamma} \in \mathbb{R}^N$ is randomly initialized and then tuned alone by optimizing a Poisson log-likelihood with a rate as $\log \lambda_{ij}=\gamma_i+\gamma_j$. Next, the latent embeddings matrix $\mathbf{W}$ is initialized based on the eigenvalues obtained by the spectral decomposition of the normalized Laplacian matrix of the network \cite{10.5555/2980539.2980649,868688}. In all experiments, we compare against unsupervised methods, and we do not include GNNs since they perform poorly in unsupervised tasks due to the over-smoothing effect \cite{gnn_bad}.

\begin{table}[!t]
\centering
\caption{Network statistics; $|\mathcal{V}|$: \# Nodes, $ |\mathcal{E}|$: \# Edges, $|\mathcal{K}|$: \# Communities.}
\label{tab:network_statistics}
\resizebox{1\textwidth}{!}{%
 \begin{tabular}{rcccccccc}\toprule
 & \textsl{AstroPh}\cite{snapnets}  & \textsl{GrQc}\cite{snapnets} & \textsl{Facebook}\cite{snapnets} & \textsl{HepTh}\cite{snapnets}& \textsl{Hamilton}\cite{fb_nets}  & \textsl{Amherst}\cite{fb_nets} & \textsl{Rochester}\cite{fb_nets} & \textsl{Mich}\cite{fb_nets} \\\midrule
$|\mathcal{V}|$ & 17,903 & 5,242 & 4,039 & 8,638 & 2,118 & 2,021 & 4,145 & 2,933 \\
$|\mathcal{E}|$ & 197,031 & 14,496 & 88,234 & 24,827 & 87,486 & 87,496 & 145,305 & 54,903 \\
$|\mathcal{K}|$& - & - & - & - & 15 & 15 & 19 & 13  \\\bottomrule
\end{tabular}%
}
\end{table}

\textbf{Link prediction:}
For the link prediction experiments, we follow the well-established strategy \cite{deepwalk-perozzi14, node2vec-kdd16} and remove $50\%$ of the network edges while keeping the residual network connected. The removed edges combined with a sample of the same number of node pairs (which are not the edges of the original network) construct the negative instances for the testing set. We utilize the residual network to learn the node embeddings. 

We consider four networks with unknown community structures and asses performance across different dimensions. In Table \ref{tab:auc_roc}, we compare the results of our method with other prominent GRL and NMF approaches in terms of the Area Under Curve-Receiver Operating Characteristic (AUC-ROC). All baselines have been tuned and feature vectors for dyads are constructed based on binary operators (average, Hadamard, weighted-L1, weighted-L2) \cite{node2vec-kdd16}. For these constructed feature vectors we further train a logistic regression model with $L_2$ regularization to make predictions. In particular, for the baselines we choose the hyperparameter settings for each model, as well as, the binary operator for which the logistic regression predictions return the maximum AUC-ROC score. 

In contrast, for our models we adopt an unbiased evaluation, and we choose the first of the considered $\delta$ values which keeps the solution identifiable (at least one champion per community), as $\delta$ decreases. We note though, the existence of identifiable regimes with higher predictive power. Furthermore, predictions and AUC-ROC scores for \textsc{HM-LDM}, can be obtained directly (without the use of a logistic regression model) and are based on the learned Poisson rates $\lambda_{ij}$ of the test set pairs $\{i,j\}$. The true dimensions for \textsc{\modelabbrv} are $D+1$ but reported as $D$ since they express the true number of model parameters, for a fair comparison with the baselines. For our method, we show the mean performance over five independent runs (error bars were found to be in the scale of $10^{-3}$ and thus not presented). 

Comparing now the results with the non-NMF models, we observe that our \textsc{\modelabbrv} (either $p=1$ or $p=2$) outperforms the baselines and in most cases significantly, returning favorable results. For the NMF models, we see mostly a big performance gap with the \textsc{\modelabbrv}, showcasing the existence of regimes for $\delta$ where we can successfully achieve identifiable community memberships while also exhibiting the link prediction power of the \textsc{LDM}. (AUC Precision-Recall scores are similar to the AUC-ROC scores and thus not presented)

\begin{table*}[!t]
\centering
\caption{Area Under Curve (AUC-ROC) scores for varying representation sizes.}
\label{tab:auc_roc}
\resizebox{0.85\textwidth}{!}{%
\begin{tabular}{rcccccccccccc}\toprule
\multicolumn{1}{l}{} & \multicolumn{3}{c}{\textsl{AstroPh}} & \multicolumn{3}{c}{\textsl{GrQc}} & \multicolumn{3}{c}{\textsl{Facebook}}& \multicolumn{3}{c}{\textsl{HepTh}}\\\cmidrule(rl){2-4}\cmidrule(rl){5-7}\cmidrule(rl){8-10}\cmidrule(rl){11-13}
\multicolumn{1}{r}{Dimension ($D$)} & $8$ & $16$ & $32$ & $8$ & $16$ & $32$ & $8$ & $16$ & $32$& $8$ & $16$ & $32$ \\\cmidrule(rl){1-1}\cmidrule(rl){2-2}\cmidrule(rl){3-3}\cmidrule(rl){4-4}\cmidrule(rl){5-5}\cmidrule(rl){6-6}\cmidrule(rl){7-7}\cmidrule(rl){8-8}\cmidrule(rl){9-9}\cmidrule(rl){10-10}\cmidrule(rl){11-11}\cmidrule(rl){12-12}\cmidrule(rl){13-13}
\textsc{DeepWalk}\cite{deepwalk-perozzi14}    &.945 	&.950	&.952  & .919	&.916 &	.929 &   .986 &	.986	& .984 &.874 &.867 & .873 \\
\textsc{Node2Vec}\cite{node2vec-kdd16}    &.950 	&\underline{.962}	&\underline{.957} & .897	&.913	&.930    & \underline{.988}    & \underline{.988} & \underline{.987} &.881 &.882 &.881   \\
\textsc{LINE}  \cite{line}      &.909 	&.938	&.947 & .920 &.925	&.919 & .981 & .987 & .983 &.873 &.886 &.882 \\
\textsc{NetMF} \cite{netmf-wsdm18}     &.813 	&.823	&.839 & .860 & .866&.877	 &  .935 & .963   & .971  &.792&.806 &.821\\
\textsc{NetSMF} \cite{netsmf-www2019}     & .891	&.901	&.919 & .837&.858	&.886 &.975  &.981  &.985 & .809&.822  &.836  \\
\textsc{LouvainNE}\cite{louvainNE-wsdm20}   & .813	&.811	&.819 & .868	&.875	&.873 & .958 &.961 &.963 &.874 &.867 &.873 \\
\textsc{ProNE}\cite{prone-ijai19}  & .907	&.929	&.947 & .885	&.911	&.921 & .971 & .982 & .987   &.827 &.846 &.859  \\\midrule
\textsc{NNSED}\cite{NNSED}   &.861 	&.882	&.891   & .792 	&.808 	&.828    &.908   &.927   &.935    & .756   & .779   &.796  \\
\textsc{MNMF}\cite{MNMF}   & .893	&.925	&.943   & .911 	&.928 	&.937    &.965   &.978   &.982    & .857   &.880    &.891  \\
\textsc{BigClam}\cite{nmf3}  &.500 	&.723	&.810   & .752 	&.769 	&.780    & .744  &.722   &.647    & .776   &.700    &.748  \\
\textsc{SymmNMF}\cite{SymmNMF}  &.767 	&.779	&.800   & .729 	&.772 	&.835    & .933  &.942   &.951    & .696   &.727    &.766  \\\midrule
\textsc{HM-LDM($p=1$)}  & \underline{.956}	&.952	&.952   &\textbf{.944} 	&\textbf{.948}	&\textbf{.951}   & .982  & .979 & .974   &\textbf{.916}   & \textbf{.921}  &\textbf{.924}
\\
\textsc{\modelabbrv ($p=2$)}      &\textbf{.972}   &\textbf{.973}   & \textbf{.963}  &\underline{.940} & \underline{.942} & \underline{.946} & \textbf{.992}     & \textbf{.993}      & \textbf{.993}      &\underline{.908} &\underline{.910} &\underline{.911}
\\\bottomrule    
\end{tabular}%
 }
\end{table*}
 
\begin{figure}[!b]
  \centering
  \subfloat[\textsl{AstroPh}]{{\includegraphics[width=0.23\columnwidth]{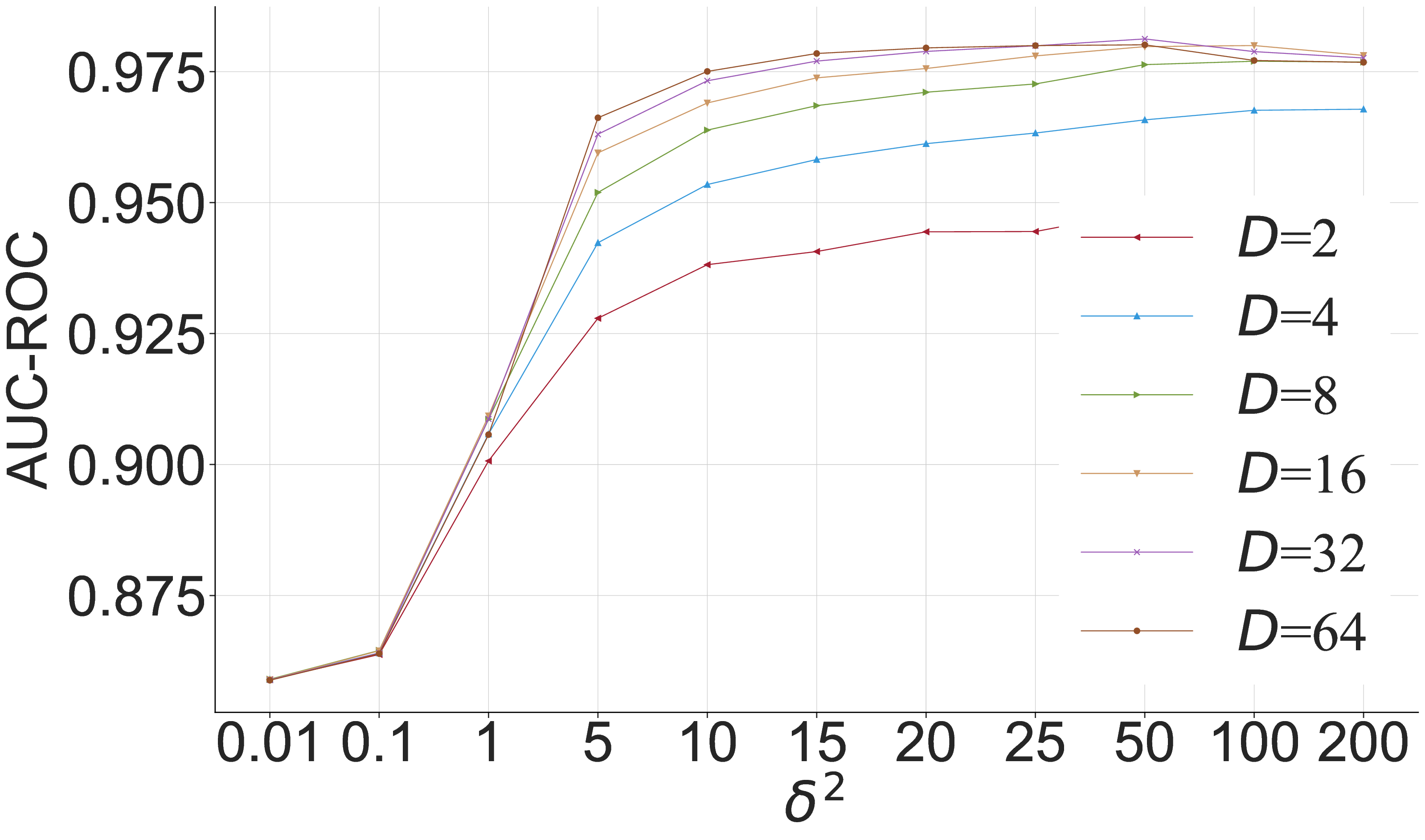} }}%
\hfill 
  \subfloat[\textsl{Facebook}]{{ \includegraphics[width=0.23\columnwidth]{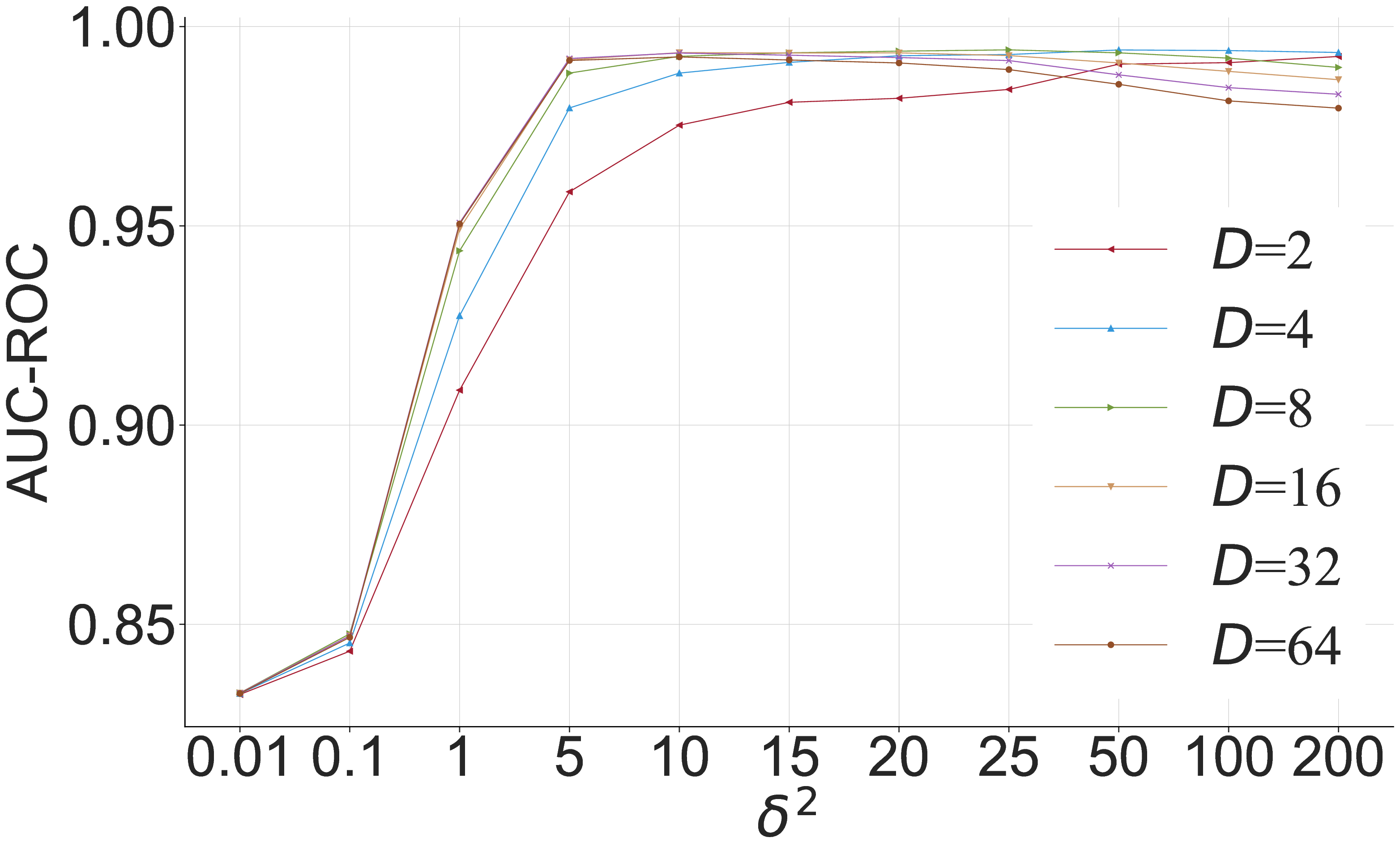} }}%
\hfill 
  \subfloat[\textsl{GrQc}]{{ \includegraphics[width=0.23\columnwidth]{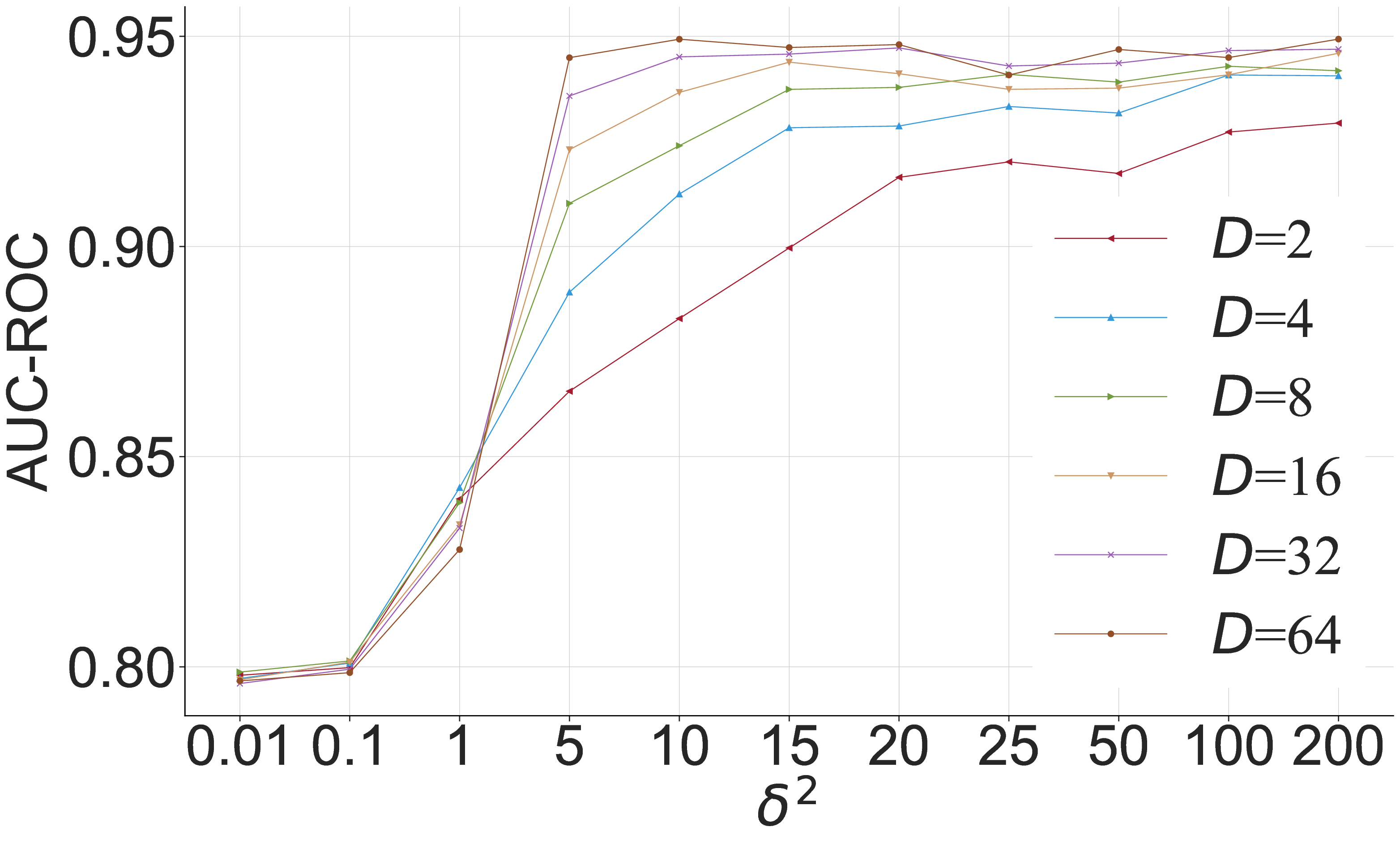} }}%
\hfill
  \subfloat[\textsl{HepTh}]{{ \includegraphics[width=0.23\columnwidth]{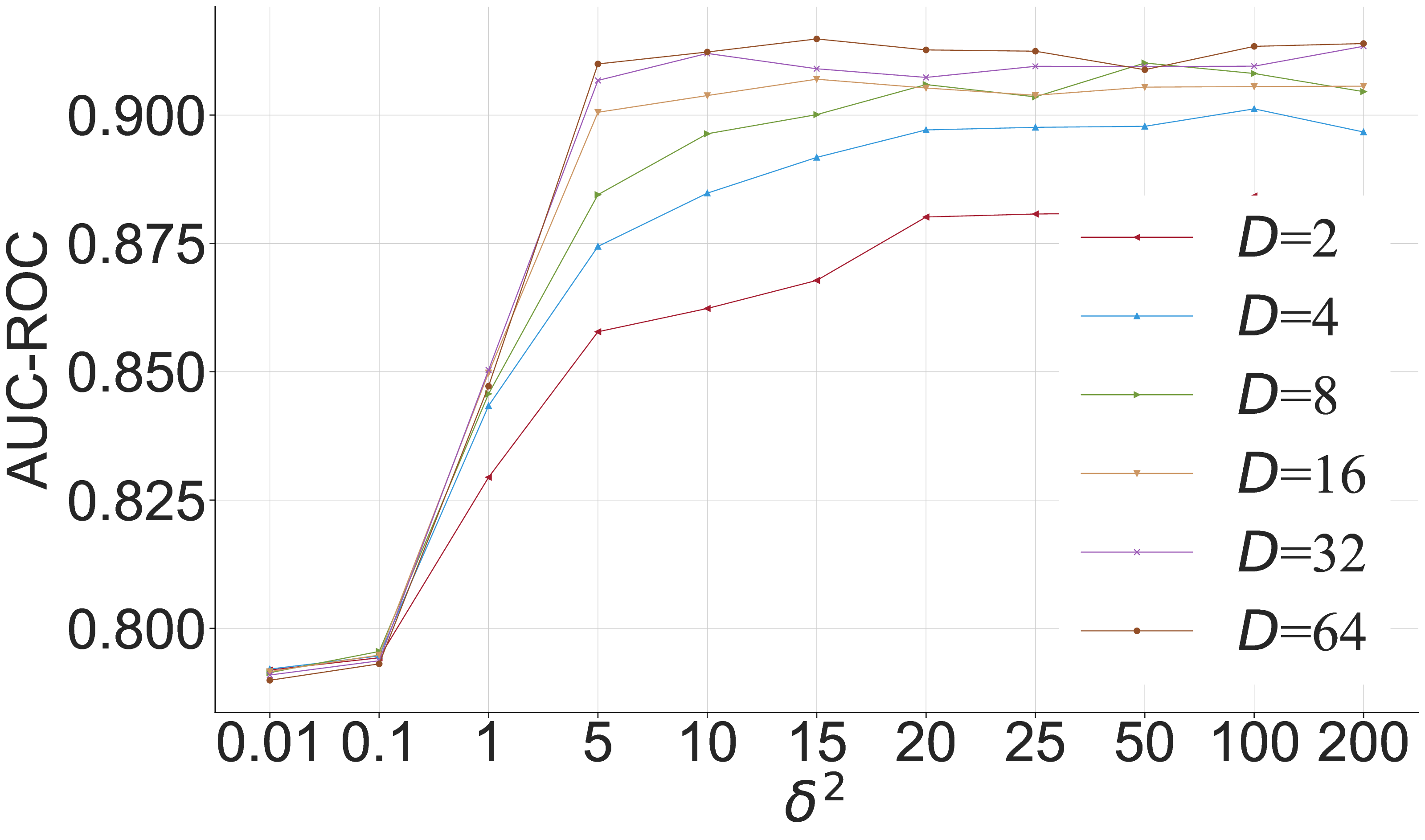} }}%
  \hfill
    \subfloat[\textsl{AstroPh}]{{ \includegraphics[width=0.23\columnwidth]{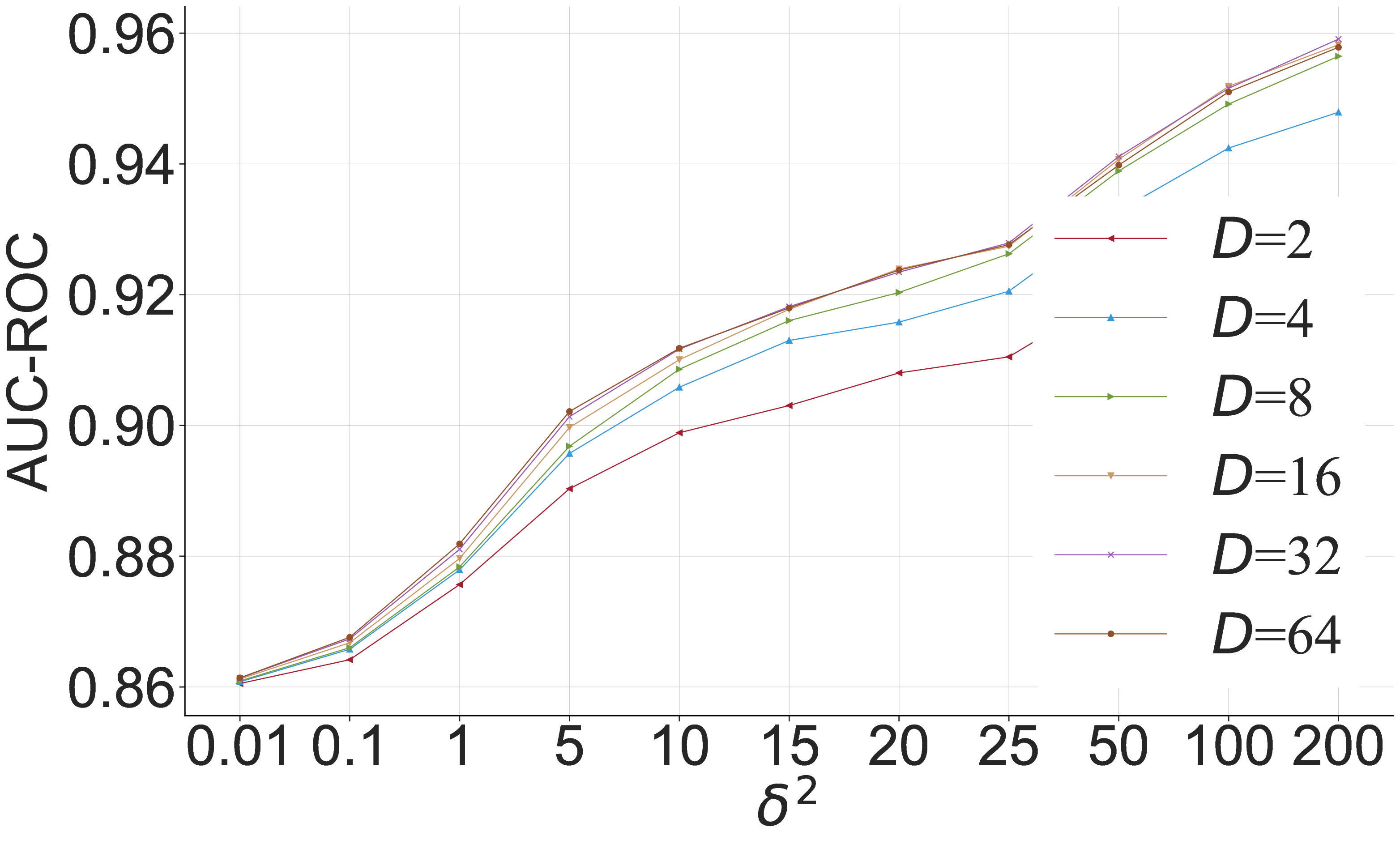} }}%
\hfill 
  \subfloat[\textsl{Facebook}]{{ \includegraphics[width=0.23\columnwidth]{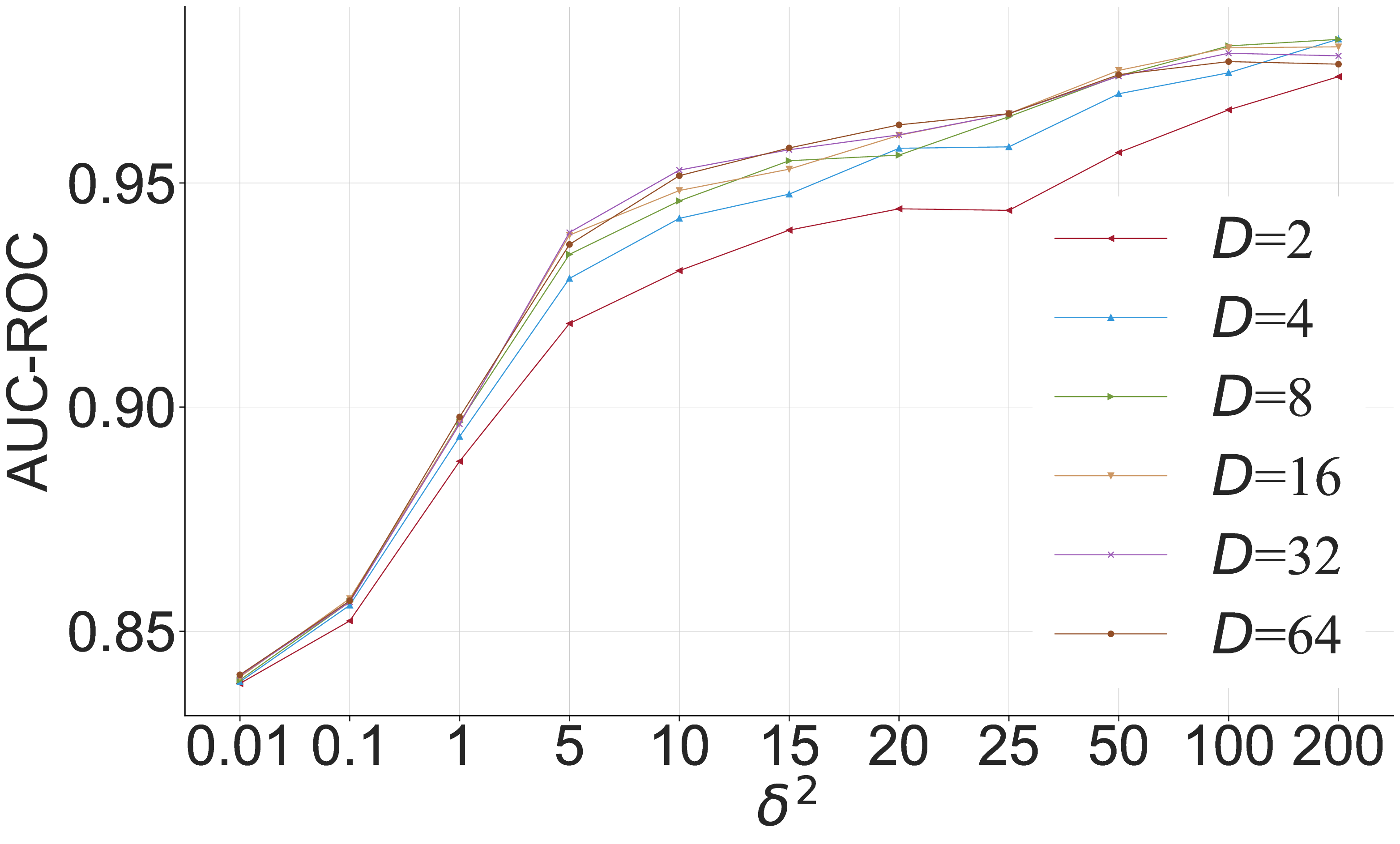} }}%
\hfill 
  \subfloat[\textsl{GrQc}]{{ \includegraphics[width=0.23\columnwidth]{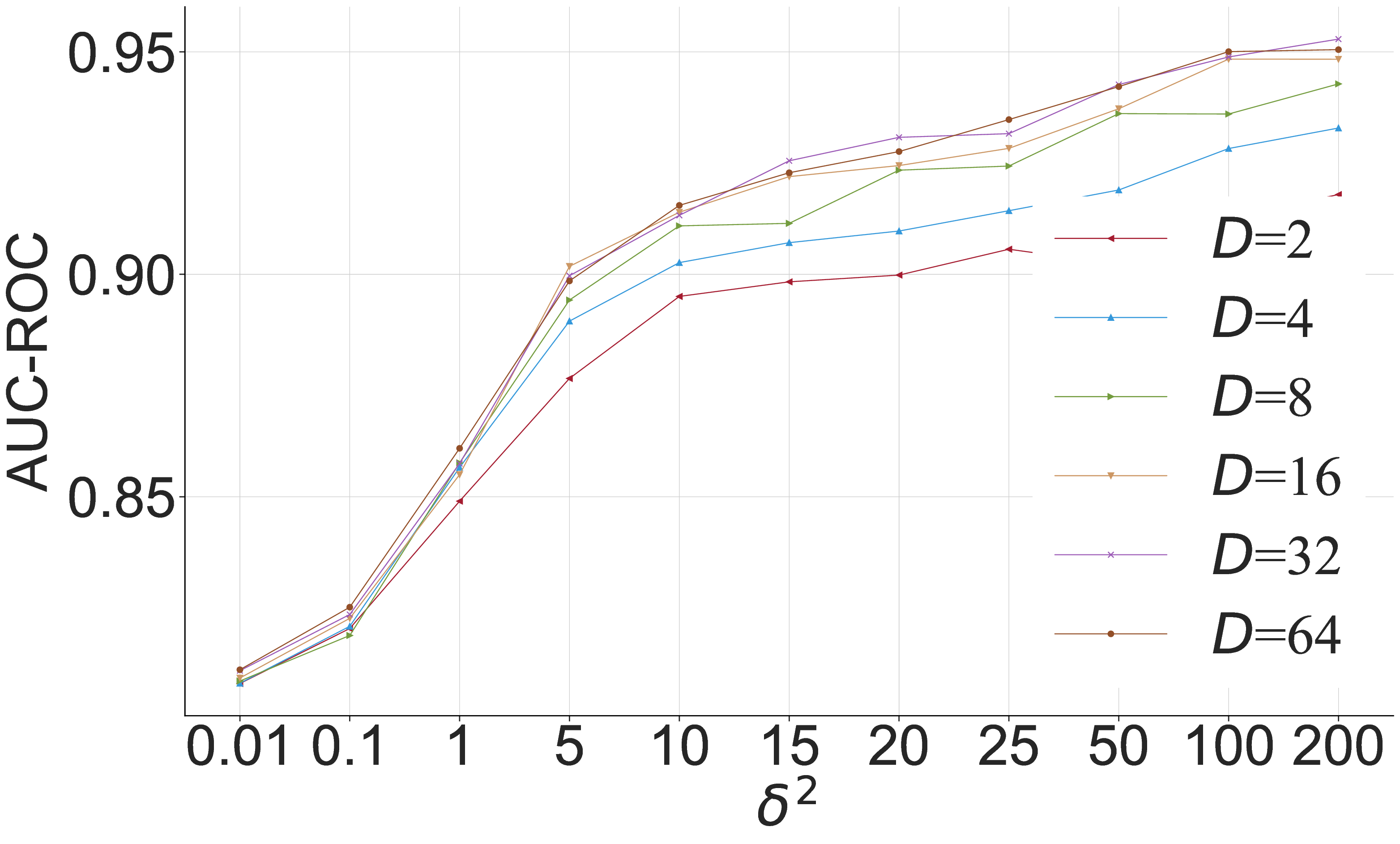} }}%
\hfill
  \subfloat[\textsl{HepTh}]{{ \includegraphics[width=0.23\columnwidth]{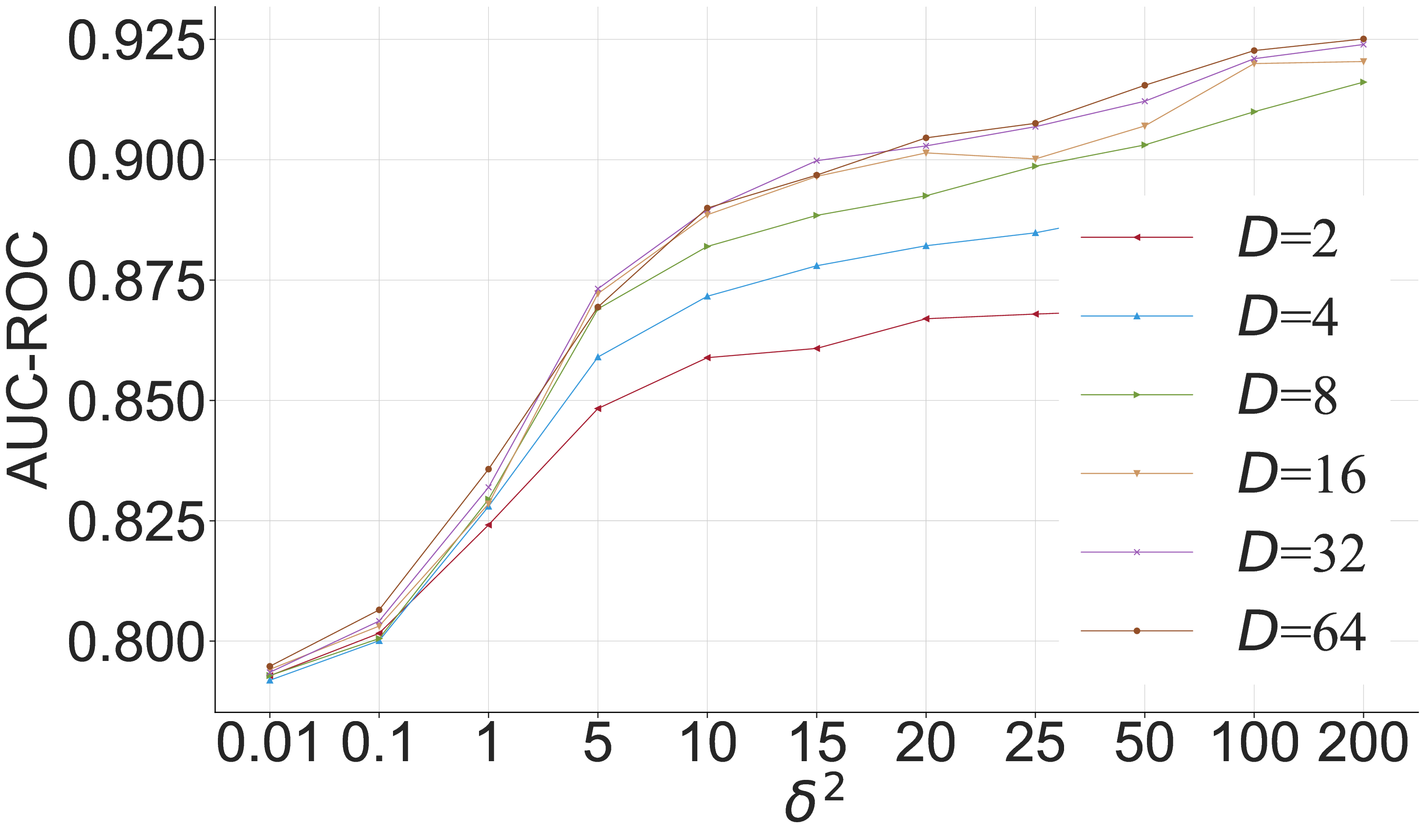} }}%
  \caption{AUC-ROC scores as a function of $\delta^2$ across dimensions for \textsc{\modelabbrv}. Top row: $p=2$. Bottom row $p=1$.}\label{fig:roc_d}
\end{figure}

 \textbf{Performance and simplex sizes:} In Figure \ref{fig:roc_d}, we provide the link prediction performance as a function of $\delta^2$ in terms of the AUC-ROC scores across various latent dimensions, networks and for both $p=1$ and $p=2$. We here observe that small $\delta$ values provide the minimum scores. This phenomenon is anticipated due to the fact that homophily properties are not sufficiently met (except within clusters) due to the very small simplex volume that these low $\delta$ values define. Rethinking \textsc{\modelabbrv} with $p=2$ as a positive Eigenmodel, we can also notice how the positivity constraint on the $\Lambda$ diagonal matrix does not allow for stochastic equivalence properties which would essentially boost performance even on low simplex volumes. As we increase the values of $\delta$, we naturally reach the performance of an unconstrained \textsc{LDM}. Comparing now, the squared and simple $\ell^2$-norm metric we observe that the former converges to performance saturation more rapidly.
 
\begin{figure} [!t]
  \centering
  \subfloat[\textsl{AstroPh}]{{ \includegraphics[width=0.23\columnwidth]{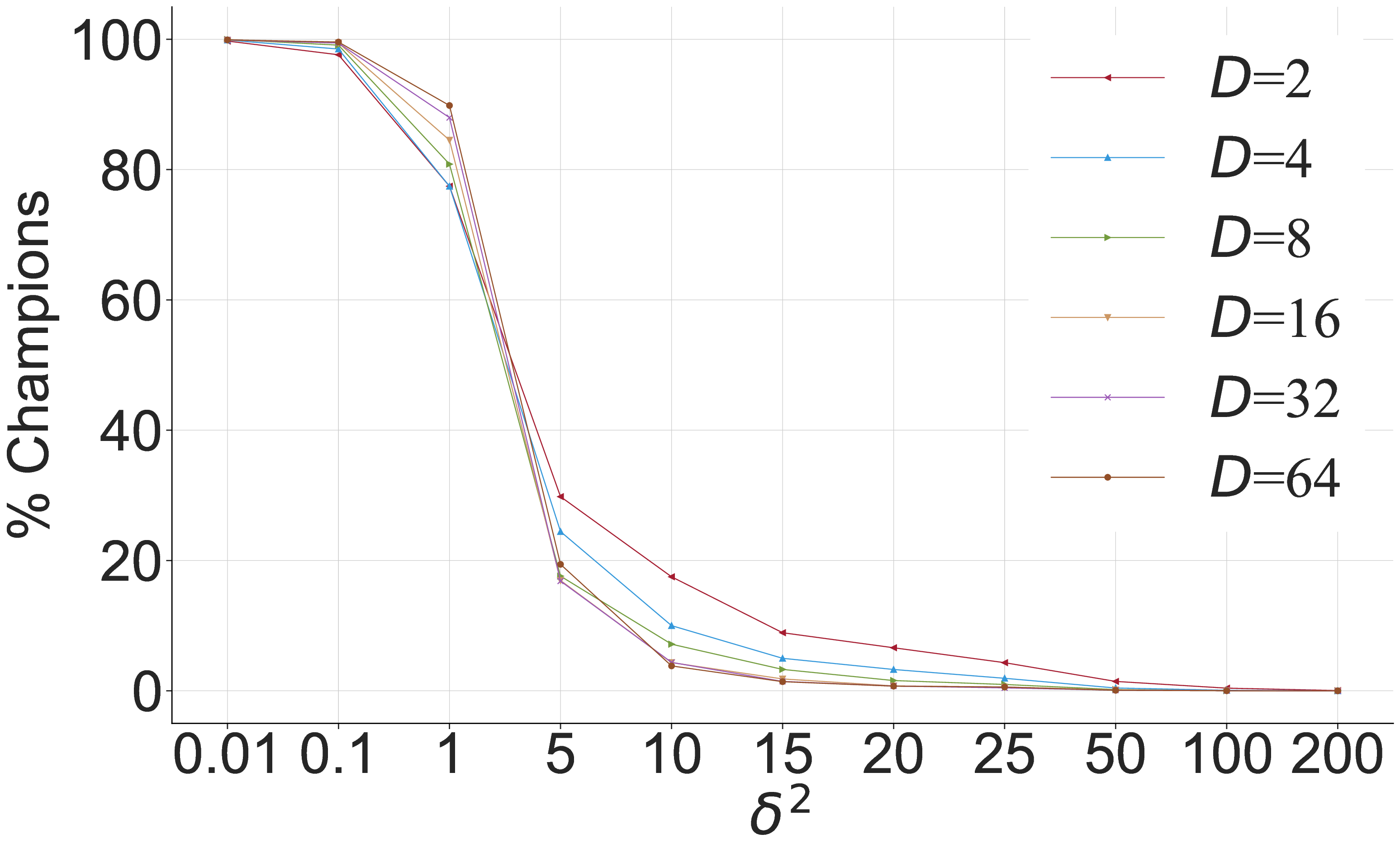} }}%
  \hfill
  \subfloat[\textsl{Facebook}]{{ \includegraphics[width=0.23\columnwidth]{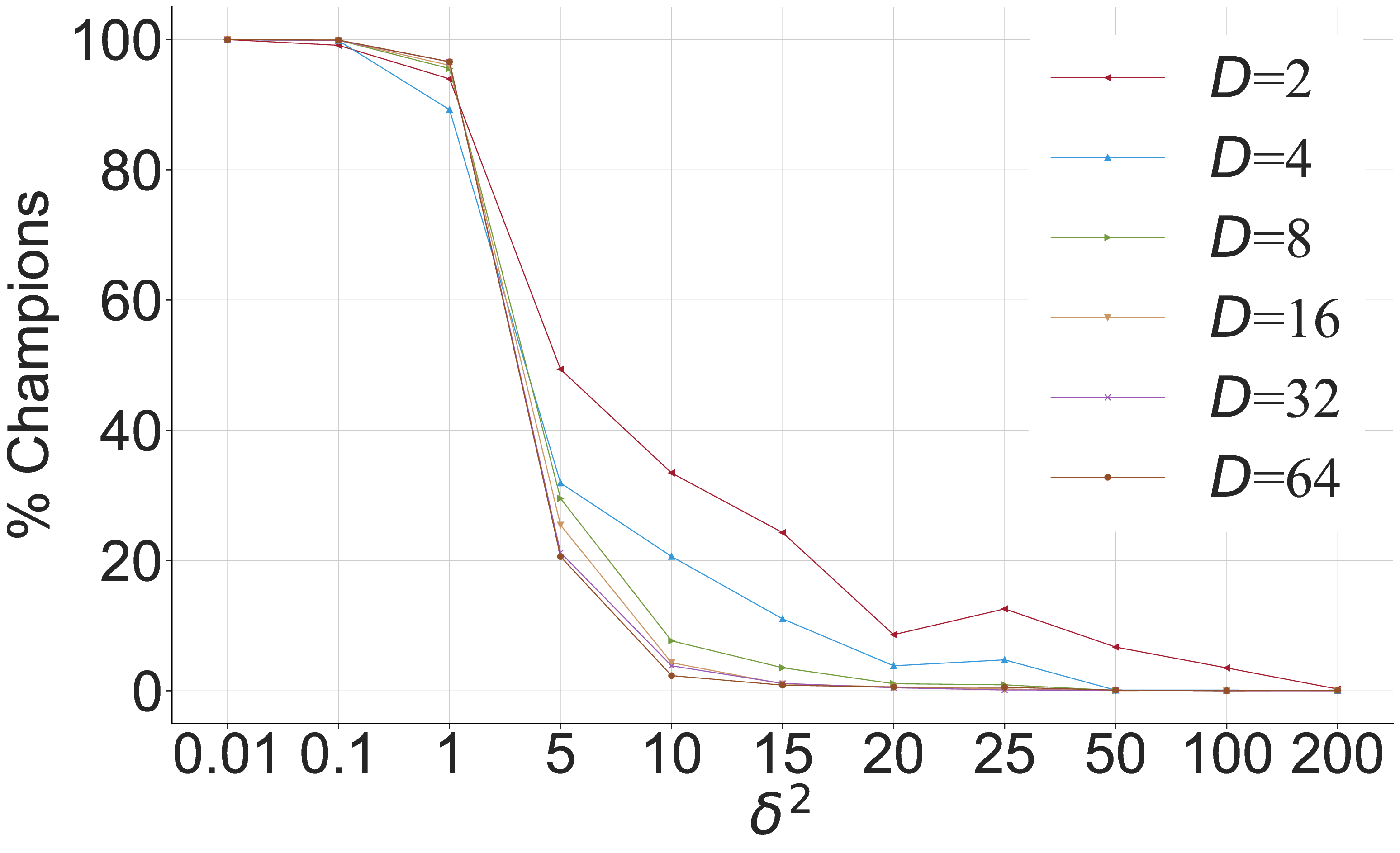} }}%
  \hfill
  \subfloat[\textsl{GrQc}]{{ \includegraphics[width=0.23\columnwidth]{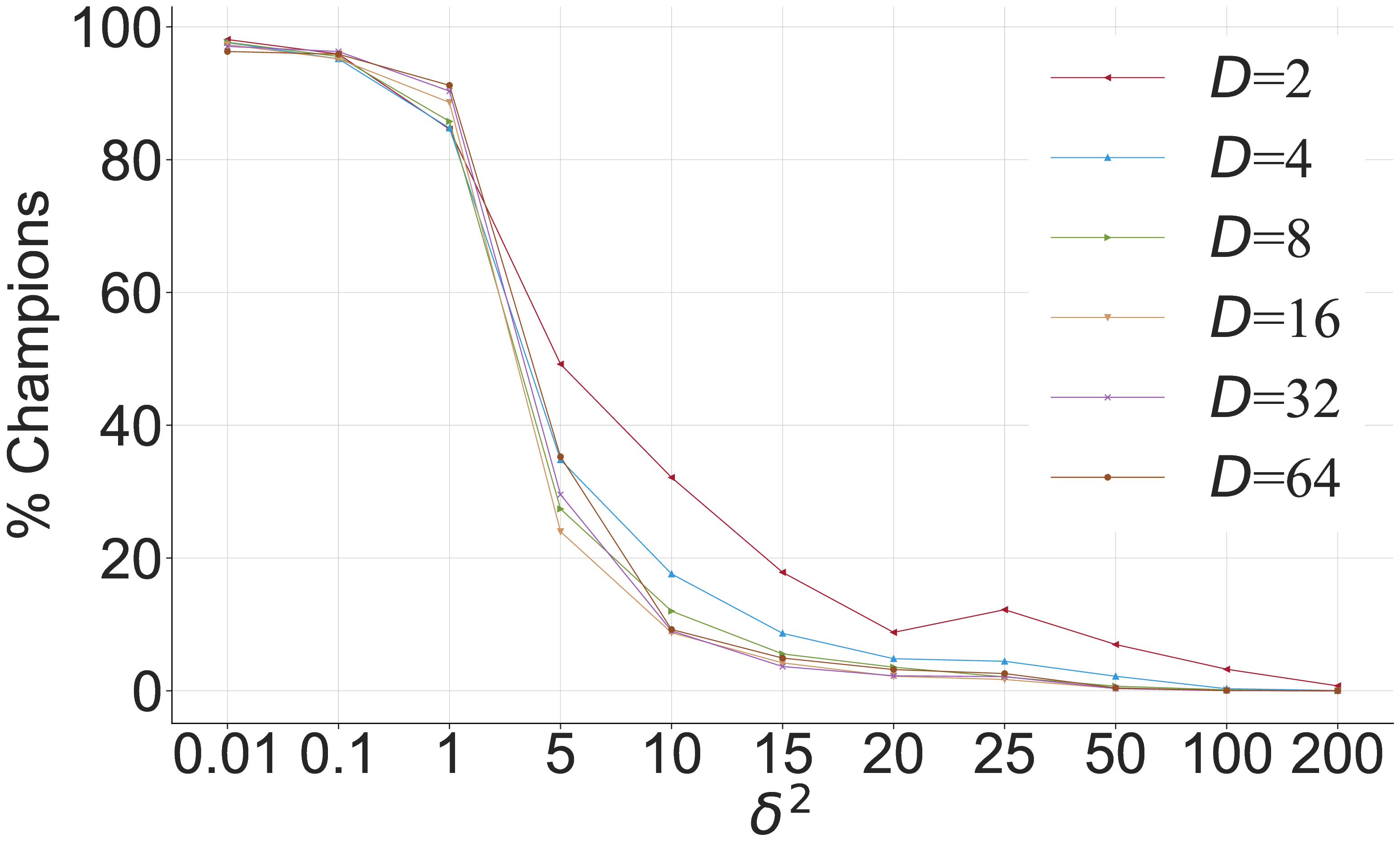} }}%
  \hfill
  \subfloat[\textsl{HepTh}]{{ \includegraphics[width=0.23\columnwidth]{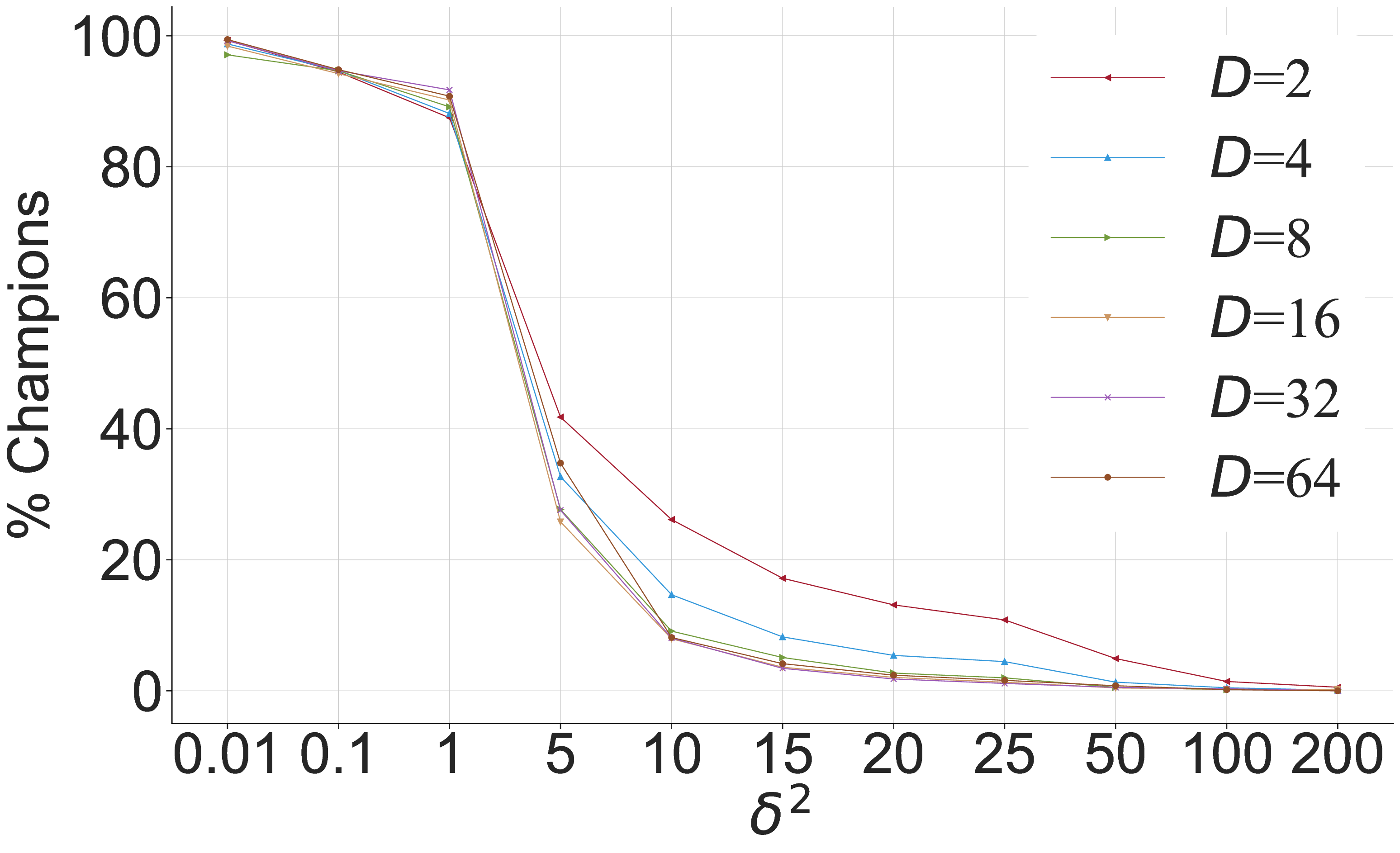} }}%
  \hfill
    \subfloat[\textsl{AstroPh}]{{ \includegraphics[width=0.23\columnwidth]{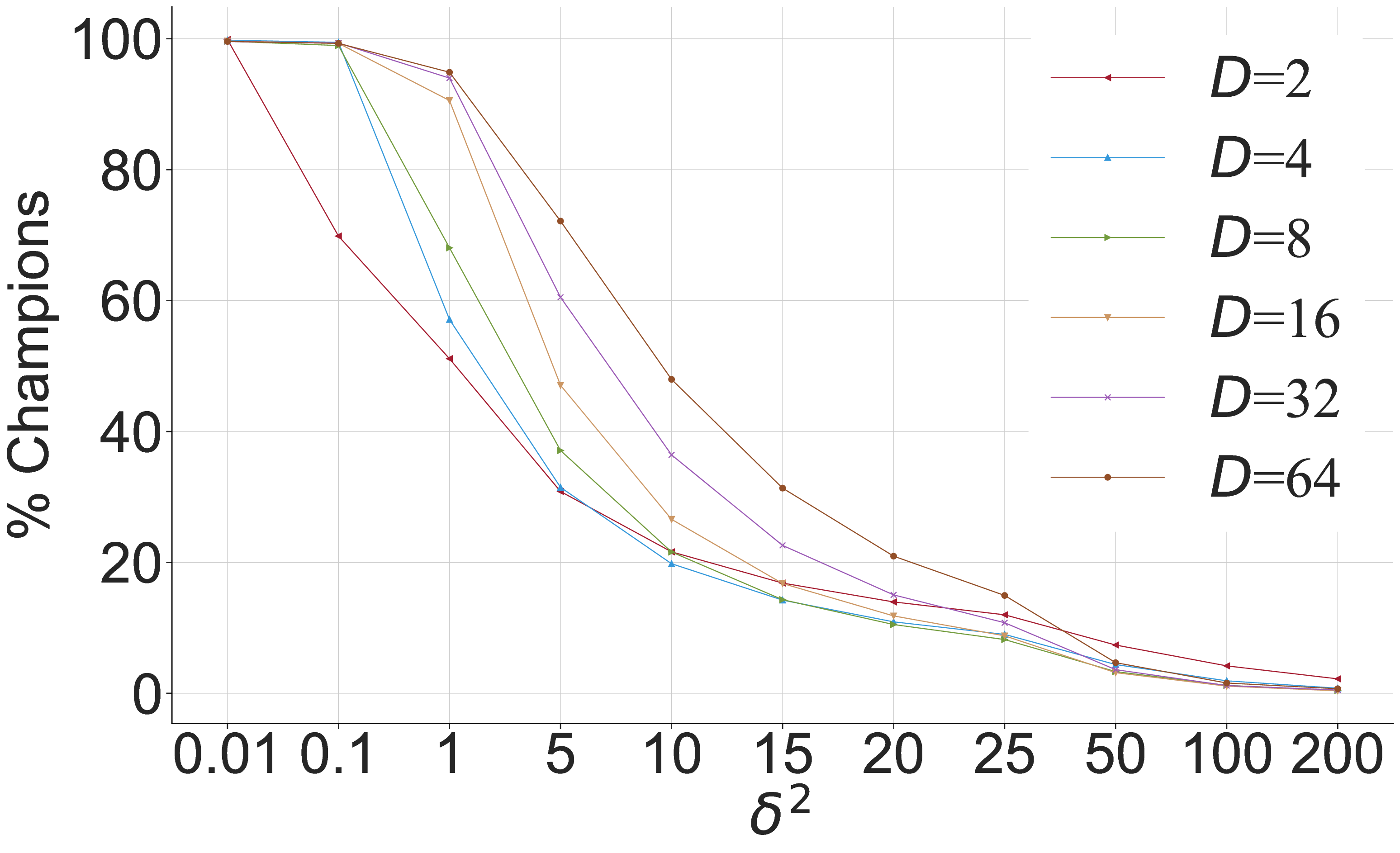} }}%
  \hfill
  \subfloat[\textsl{Facebook}]{{ \includegraphics[width=0.23\columnwidth]{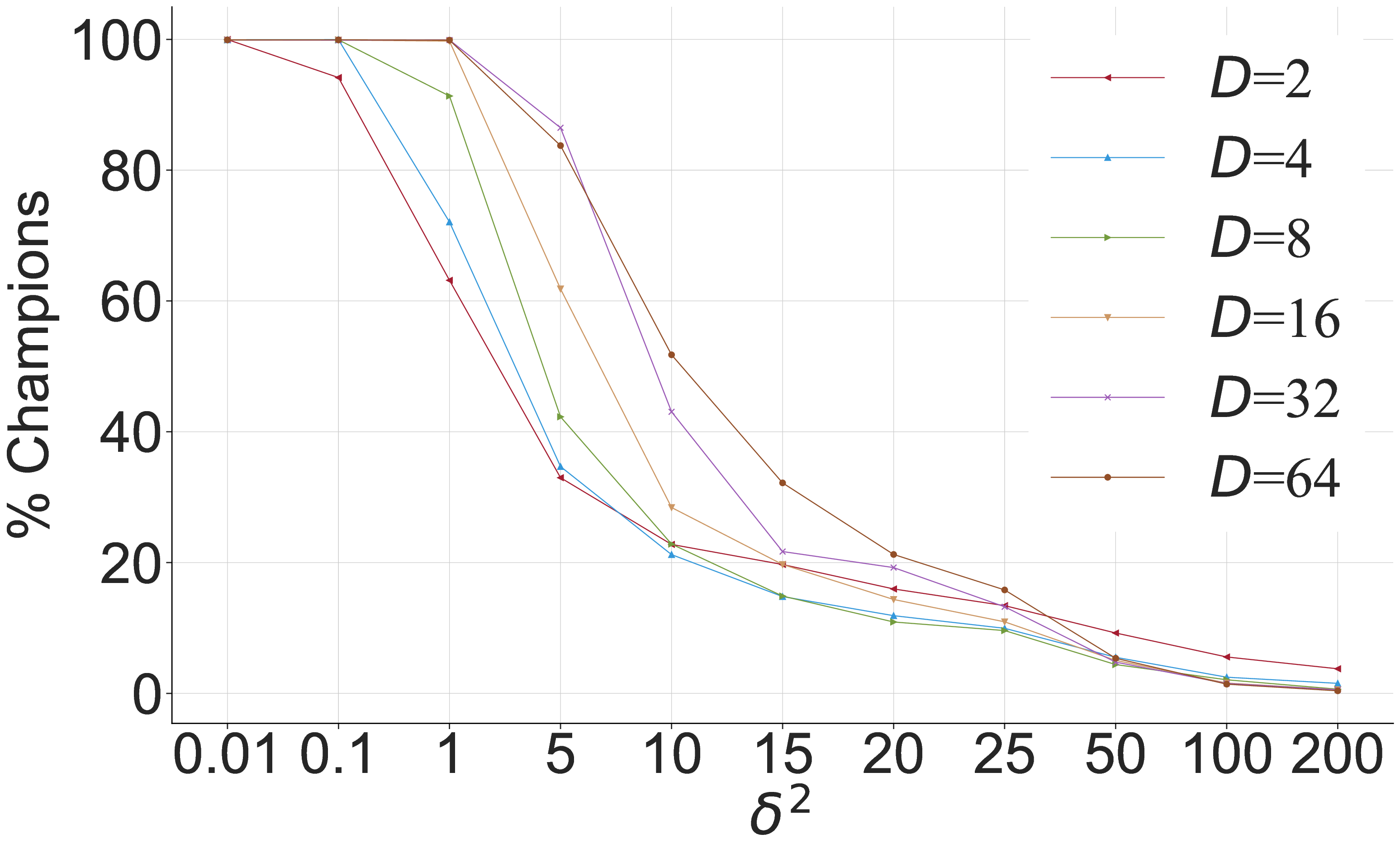} }}%
  \hfill
  \subfloat[\textsl{GrQc}]{{ \includegraphics[width=0.23\columnwidth]{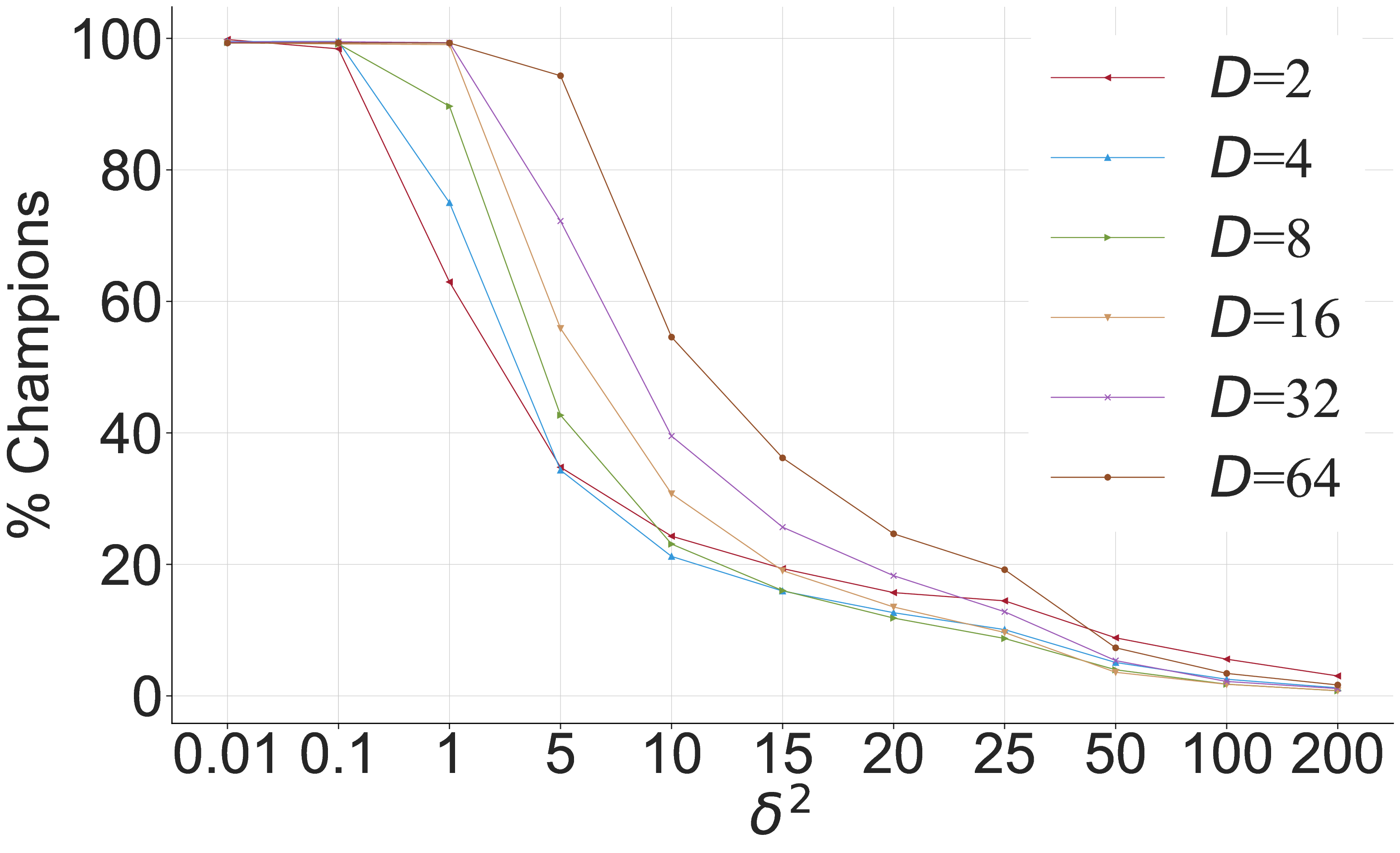} }}%
  \hfill
  \subfloat[\textsl{HepTh}]{{ \includegraphics[width=0.23\columnwidth]{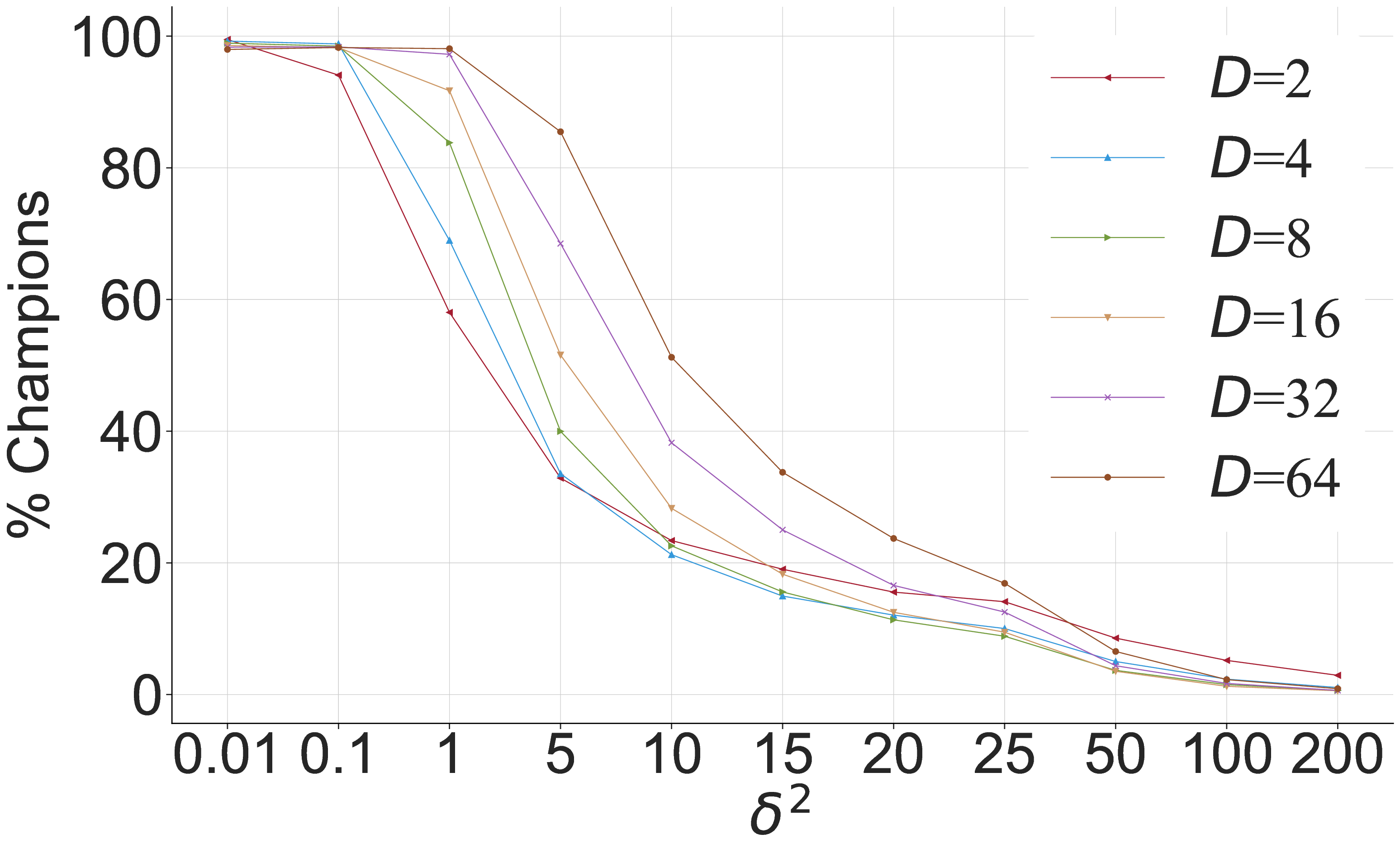} }}%
  \caption{Total community champions (\%) in terms of $\delta^2$ across dimensions for \textsc{\modelabbrv}. Top row: $p=2$. Bottom row $p=1$.}  \label{fig:phase_transitions}
\end{figure}

\begin{figure}[!t]
  \centering
  \subfloat[\textsl{GrQc} $(p=2)$]{{ \includegraphics[width=0.23\columnwidth]{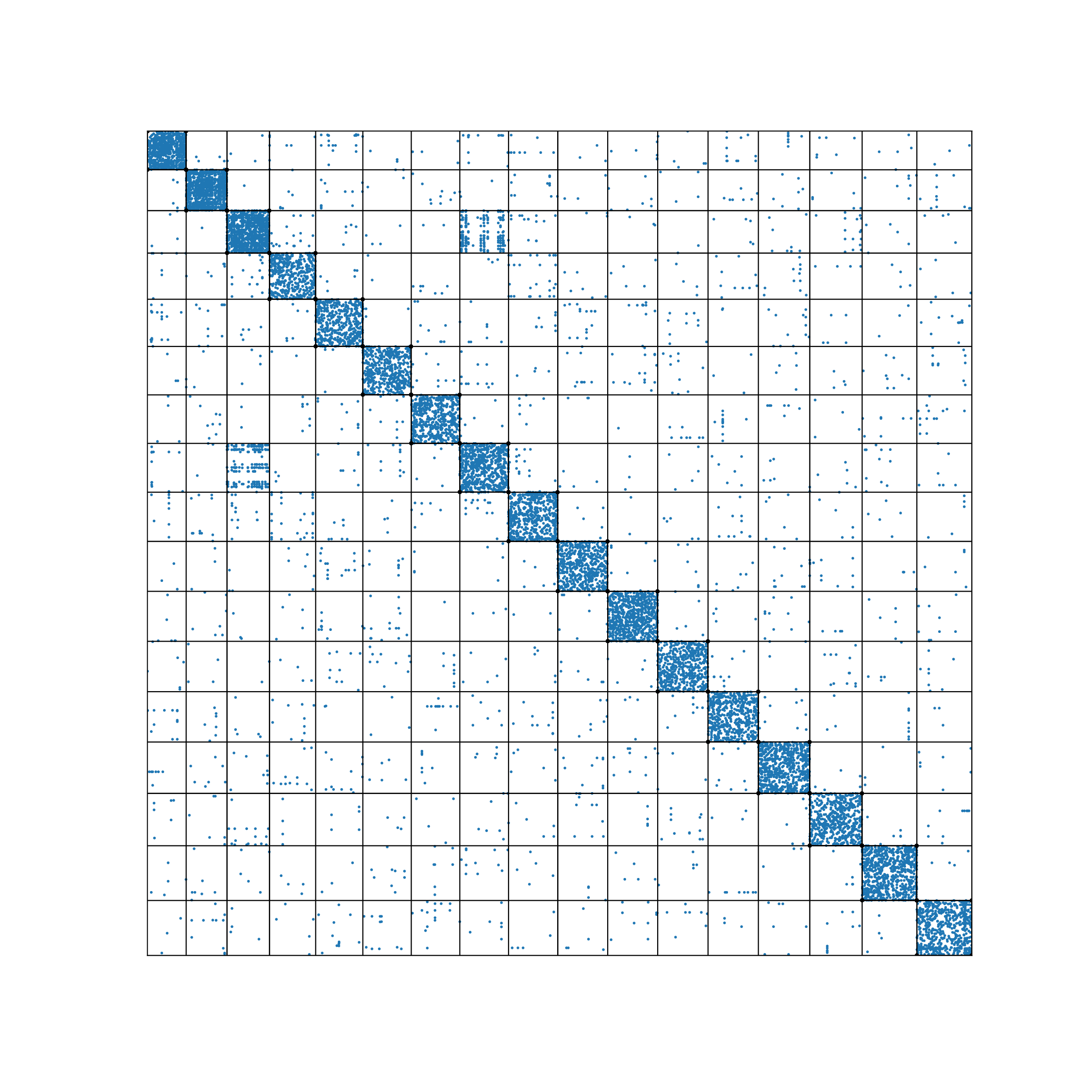} }}%
\hfill
  \subfloat[\textsl{HepTh} $(p=2)$]{{ \includegraphics[width=0.23\columnwidth]{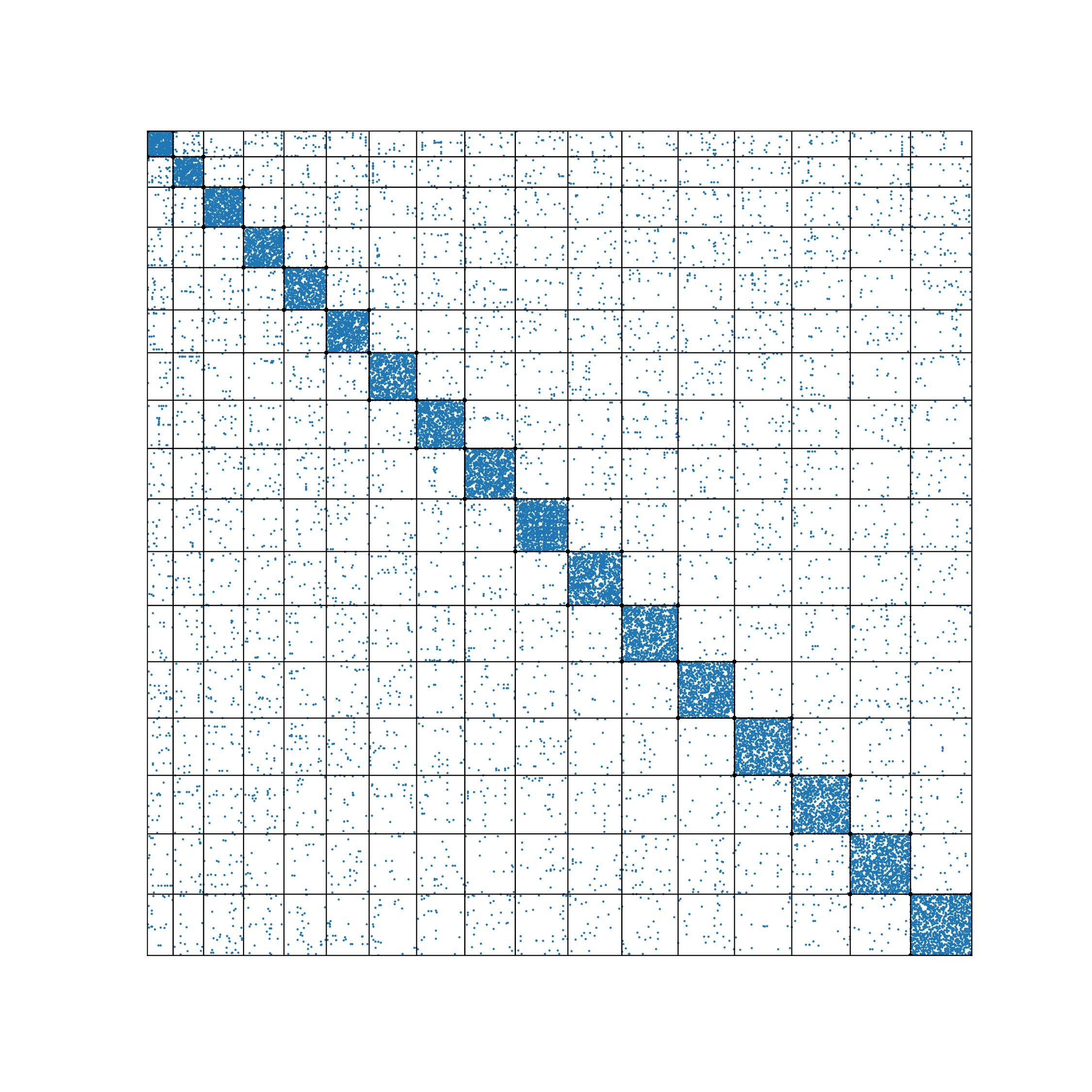} }}%
\hfill
  \subfloat[\textsl{GrQc} $(p=1)$]{{ \includegraphics[width=0.23\columnwidth]{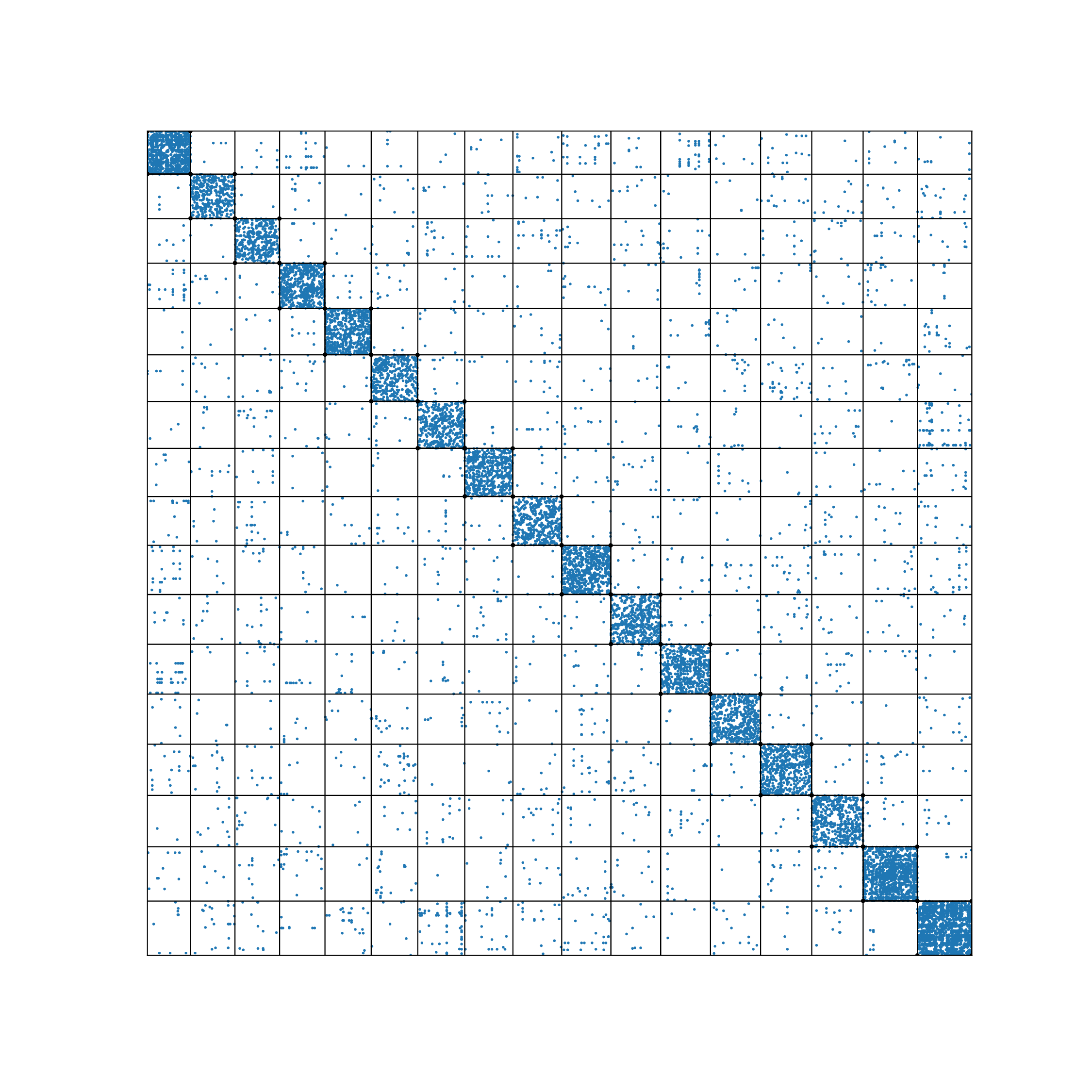} }}%
\hfill
  \subfloat[\textsl{HepTh} $(p=1)$]{{ \includegraphics[width=0.23\columnwidth]{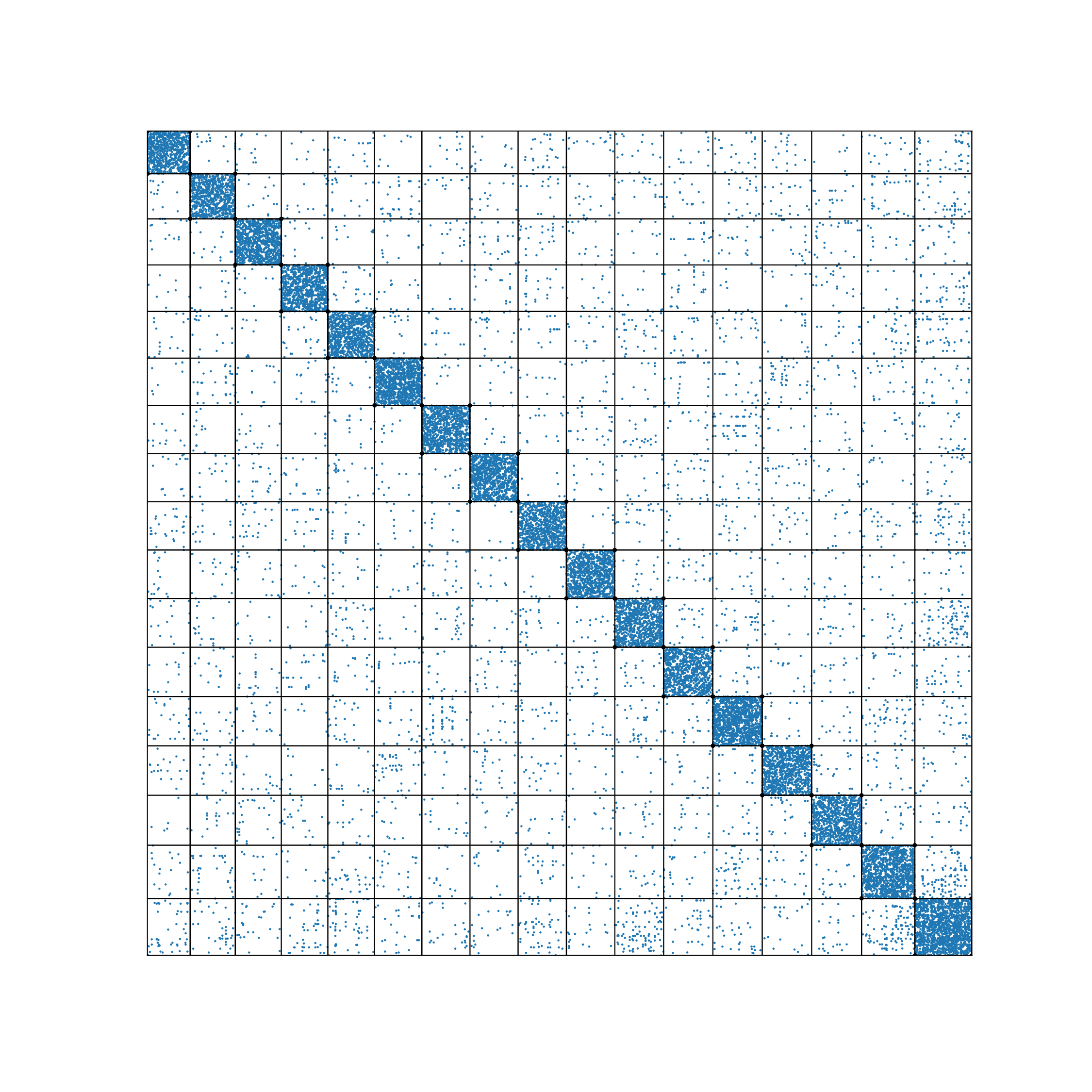} }}%
  \caption{Ordered adjacency matrices based on the memberships of a $D=16$ dimensional \textsc{\modelabbrv} with $\delta$ values ensuring identifiability.}\label{fig:adj}
\end{figure}

 \textbf{Type and quality of latent memberships:} In order to understand how the size of the simplex affects the membership types of \textsc{\modelabbrv}, we provide in Figure \ref{fig:phase_transitions} the total network percentage of community champions as a function of $\delta^2$ across various latent dimensions. As expected, for very small values of $\delta$ almost $100\%$ of nodes are assigned solely to a unique simplex corner, yielding hard cluster assignments. As we increase $\delta$, we observe that more and more nodes are assigned with mixed-memberships; on the other hand, the number of champions goes to zero across all dimensions for large values of $\delta$. Contrasting again, the different powers $p$ of the \textsc{\modelabbrv} formulation, we notice that the decrease in community champions is steeper for $p=2$. This also explains why the squared $\ell^2$ choice leads to faster convergence in the AUC-ROC, as the model converges faster to the classic \textsc{LDM}. Overall, it is evident that the $p=2$ \textsc{\modelabbrv} needs smaller simplex volumes to be identifiable. We continue with assessing unique latent structures of \textsc{\modelabbrv}. For that purpose, in Figure \ref{fig:adj} we provide the reorganized adjacency matrices with respect to the community allocations of \textsc{\modelabbrv} (for mixed-memberships we assign a node based on the maximum membership). We witness how \textsc{\modelabbrv} successfully discovers latent communities, facilitating part-based network representations while choosing appropriate $\delta$ regimes ensure identifiability.

 \textbf{Experiments using real ground-truth communities:} In order to assess the ability of \textsc{HM-LDM} to discover informative communities, we make use of four networks providing ground-truth community labels. For the NMF-based methods, including ours, we test the ability of the algorithms to detect valid structures by comparing the inferred memberships with the ground-truth community labels while we set the latent dimensions to be equal to the total number of communities. For the GRL approaches which do not define memberships, we extract latent embeddings and use {\it k-means} (average over 20 runs for robustness) to obtain memberships. We report the Normalized Mutual Information (NMI) score, as well as, the Adjusted Rand Index (ARI), both measures have been validated for community quality assessment in \cite{com_metrics}. Again, all the baselines have been tuned individually for each network in terms of their hyperparameters. In contrast, for our \textsc{HM-LDM}, we do not perform any tuning and we just set $\delta=1$ for all networks since this choice provides in general informative and mostly hard cluster assignments. For our method and the classic LDMs, we report scores averaged over five independent runs in each of which we run the algorithm five times extracting the model with the lowest training loss to remove the effect of local-minimas. We summarize our findings in Table \ref{tab:nmi_ari}, where we witness mostly favorable or on-par performance of \textsc{\modelabbrv} with all of the competitive baselines for the NMI metric. For the ARI metric, we observe that our framework outperforms significantly the baselines in all of the considered networks.

\begin{table*}[!t]
\centering
\caption{Normalized Mutual Information (NMI) and Adjusted Rand Index (ARI) scores for networks with ground-truth communities.}
\label{tab:nmi_ari}
\resizebox{0.6\textwidth}{!}{%
\begin{tabular}{rcccccccc}\toprule
\multicolumn{1}{l}{} & \multicolumn{2}{c}{\textsl{Amherst}} & \multicolumn{2}{c}{\textsl{Rochester}} & \multicolumn{2}{c}{\textsl{Mich}}& \multicolumn{2}{c}{\textsl{Hamilton}}\\\cmidrule(rl){2-3}\cmidrule(rl){4-5}\cmidrule(rl){6-7}\cmidrule(rl){8-9}
\multicolumn{1}{c}{Metric} & NMI & ARI & NMI & ARI & NMI & ARI & NMI & ARI
\\\cmidrule(rl){1-1}\cmidrule(rl){2-2}\cmidrule(rl){3-3}\cmidrule(rl){4-4}\cmidrule(rl){5-5}\cmidrule(rl){6-6}\cmidrule(rl){7-7}\cmidrule(rl){8-8}\cmidrule(rl){9-9}
\textsc{DeepWalk}\cite{deepwalk-perozzi14}       &.498 	&.347  &.348  & .205	&.207  &.157	 & .447   &.303	  \\
\textsc{Node2Vec}\cite{node2vec-kdd16}       &.535 	&.375  & .364 & .223	& .217  &.161	 &.481   &.348	  \\
\textsc{LINE}  \cite{line}         & .549	&.452  & .365 & .217	&\textbf{.249}   &.192	 &.499  &.411		  \\
\textsc{NetMF} \cite{netmf-wsdm18}       &.491 	&.330  &.377  & .243	& .237  &.136	 &.456    &	.297	  \\
\textsc{NetSMF} \cite{netsmf-www2019}        &\underline{.562} 	&.408  &\underline{.381}  & .228	& \underline{.242}  &.169	 & .494   &	.391	  \\
\textsc{LouvainNE}\cite{louvainNE-wsdm20}      & \underline{.562}	&.395  &  .347 &.204	&.175  &.114 	 &.475	  &	.334  \\
\textsc{ProNE}\cite{prone-ijai19}     & .536	& .443 &.356  & .312	&.229   &\underline{.200}	 &.478    &	.396	  \\\midrule
\textsc{NNSED}\cite{NNSED}     &.295	& .243 &.168   &.116 	& .064  &.035	 & .335   &	.285 \\
\textsc{MNMF}\cite{MNMF}      & .542	&.362  & .324 & .171	&.188   &.102	 & .466   &	.287	  \\
\textsc{BigClam}\cite{nmf3}   & .091	&.066	& .028  & .022	&.024	&.015   & .053  &.041 \\
\textsc{SymmNMF}\cite{SymmNMF}  &\textbf{.596} 	&.397	& .308  &.175 	&.207	& .088  & .437  &.341   \\\midrule
\textsc{HM-LDM($p=1$)} & \underline{.562}	&\underline{.502}	&\textbf{.400}   &\textbf{.392} 	&.228	&\textbf{.205}   & \textbf{.527}  &  \underline{.485}
\\
\textsc{HM-LDM($p=2$)} &.539 	&\textbf{.506}  &\underline{.384}  & \underline{.373}	&.217 &.183	 &\underline{.507}    &	\textbf{.504}	
\\\bottomrule    
\end{tabular}%
}
\end{table*}

\textbf{Comparison with the \textsc{LDM}:} We further investigate the performance of \textsc{HM-LDM} against the LDM, including random effects for a fair comparison and for both normal and squared $\ell^2$-norm \textsc{LDM-Re} and \textsc{LDM-Re}-$(\ell^2)^2$, respectively. Towards that aim, in Table \ref{tab:auc_roc_comp} and Table \ref{tab:nmi_ari_comp} we provide the performance scores for the link prediction and clustering tasks of each model. We here witness that constraining the latent space in identifiable simplex volumes leads to a minor decrease in the predictive power, in terms of the AUC-ROC. For the community detection task, we see favorable NMI scores while the \textsc{\modelabbrv} leads to considerably higher ARI scores.  Comparing the classical LDM 
with \textsc{\modelabbrv} for $\delta^2=10^3$
provides on-par link-prediction performance but the clustering scores drop significantly. This is expected as for large simplex volumes the \textsc{\modelabbrv} approximates almost exactly the LDM with the cost of identifiability.

\begin{figure}[!b]
\centering
\subfloat[$p=1$, $\delta=1$]{{ \includegraphics[scale=0.2]{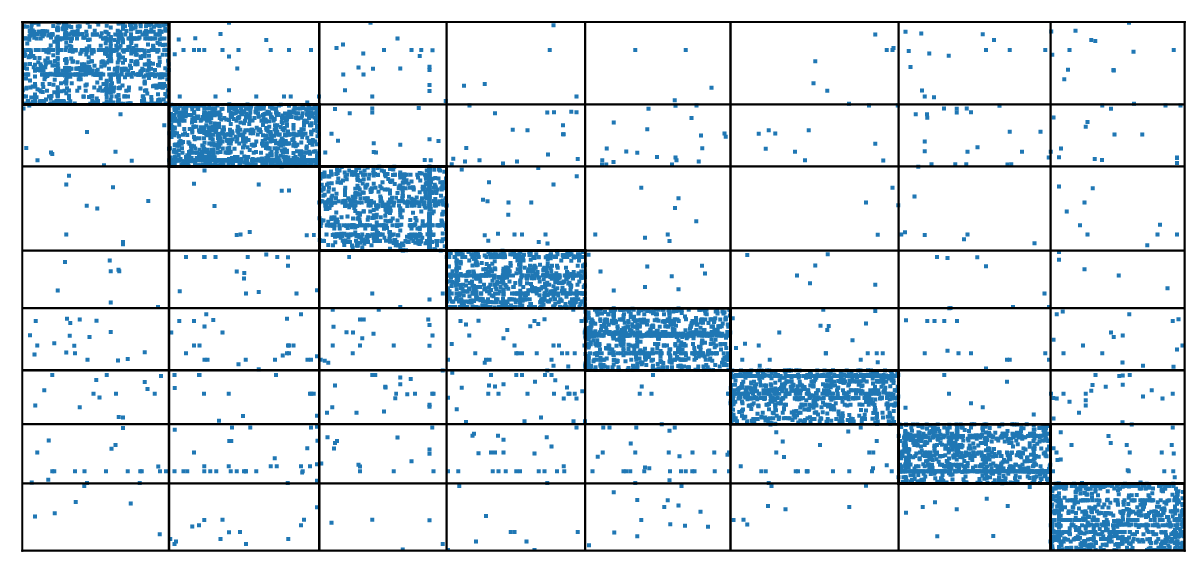} }}%
\hspace{0.5cm}
\subfloat[$p=2$, $\delta=1$]{{ \includegraphics[scale=0.2]{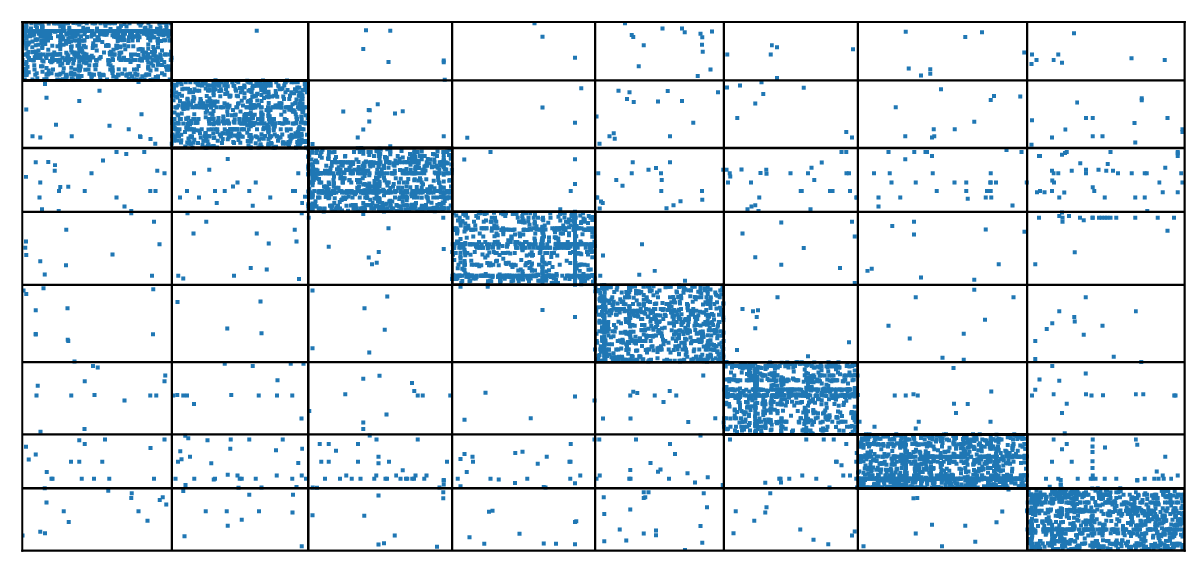} }}
\caption{\textsl{Drug-Gene} ordered adjacency matrices based on \textsc{\modelabbrv} with $D=8$.}\label{fig:bip}
\end{figure}

\textbf{Extension to bipartite networks:} Finally, we showcase the extension of our \textsc{\modelabbrv} framework to the analysis of bipartite networks. This is straightforward by introducing a different set of latent variables for the two disjoint sets of nodes, as defined by the bipartite structure. In particular, \textsc{\modelabbrv} for $p=2$,
simply extends the symmetric NMF formulation, obtained for the undirected networks, to the non-symmetric NMF specification. In Figure \ref{fig:bip}, we provide the re-ordered adjacency matrix with respect to the community allocations defined by the learned embeddings of \textsc{\modelabbrv} for a \textsl{Drug-Gene} \cite{snapnets} network ($|\mathcal{V}|=7,341|$, $|\mathcal{E}|=15,138$) where we observe a clear block structure. Importantly, the \textsc{\modelabbrv} offers identifiable joint embedding representations, mixed memberships, and community discovery for bipartite networks, tasks considered to be non-trivial and arduous.


\begin{table*}[!t]
\centering
\caption{\textsc{\modelabbrv} and \textsc{LDM-Re} comparison for the link prediction task.}
\label{tab:auc_roc_comp}
\resizebox{0.9\textwidth}{!}{%
\begin{tabular}{lcccccccccccc}\toprule
\multicolumn{1}{l}{} & \multicolumn{3}{c}{\textsl{AstroPh}} & \multicolumn{3}{c}{\textsl{GrQc}} & \multicolumn{3}{c}{\textsl{Facebook}}& \multicolumn{3}{c}{\textsl{HepTh}}\\\cmidrule(rl){2-4}\cmidrule(rl){5-7}\cmidrule(rl){8-10}\cmidrule(rl){11-13}
\multicolumn{1}{c}{Dimension ($D$)} & $8$ & $16$ & $32$ & $8$ & $16$ & $32$ & $8$ & $16$ & $32$& $8$ & $16$ & $32$ \\\cmidrule(rl){1-1}\cmidrule(rl){2-2}\cmidrule(rl){3-3}\cmidrule(rl){4-4}\cmidrule(rl){5-5}\cmidrule(rl){6-6}\cmidrule(rl){7-7}\cmidrule(rl){8-8}\cmidrule(rl){9-9}\cmidrule(rl){10-10}\cmidrule(rl){11-11}\cmidrule(rl){12-12}\cmidrule(rl){13-13}
\textsc{LDM-Re}  &.973  	&.974	&.979  & .949	&.952	&.954 & .993 & .994& .992   &.920 &.923 &.923 \\ 
\textsc{HM-LDM}($p=1,\delta^2=\text{identifiable}$)  & .956	&.952	&.952   &.944	&.948	&.951   & .982  & .979 & .974   &.916  & .921 &.924
\\
\textsc{HM-LDM}($p=1,\delta^2=10^3$) &.967 &  .967 & .965 & .956 & .955 & .951 & .985 & .986 & .987 & .932 & .931 & .926
\\\midrule
\textsc{LDM-Re}-$(\ell^2)^2$    &.979  &.978  &.976    & .944	&.944	& .945 & .990 & .990 & .991 & .913 &.912 &.909\\ 
\textsc{\modelabbrv}($p=2,\delta^2=\text{identifiable}$)      &.972   &.973  & .963 &.940 & .942 & .946 & .992   & .993     & .993     &.908 &.910 &.911
\\
\textsc{\modelabbrv}($p=2,\delta^2=10^3$) & .984& .983 &  .980 & .948 & .946 & .946 & .991 & .991 & .992 & .920 & .918 & .913
\\\bottomrule    
\end{tabular}%
 }
\end{table*}

\begin{table*}[!t]
\centering
\caption{\textsc{\modelabbrv} and \textsc{LDM-Re} comparison for the clustering task.}
\label{tab:nmi_ari_comp}
\resizebox{0.8\textwidth}{!}{%
\begin{tabular}{lcccccccc}\toprule
\multicolumn{1}{l}{} & \multicolumn{2}{c}{\textsl{Amherst}} & \multicolumn{2}{c}{\textsl{Rochester}} & \multicolumn{2}{c}{\textsl{Mich}}& \multicolumn{2}{c}{\textsl{Hamilton}}\\\cmidrule(rl){2-3}\cmidrule(rl){4-5}\cmidrule(rl){6-7}\cmidrule(rl){8-9}
\multicolumn{1}{c}{Metric} & NMI & ARI & NMI & ARI & NMI & ARI & NMI & ARI  \\\cmidrule(rl){1-1}\cmidrule(rl){2-2}\cmidrule(rl){3-3}\cmidrule(rl){4-4}\cmidrule(rl){5-5}\cmidrule(rl){6-6}\cmidrule(rl){7-7}\cmidrule(rl){8-8}\cmidrule(rl){9-9}
\textsc{LDM-Re}    & .548	&.366   &.391   &.212  	&.230    &.132	 & .491  & .320	  \\ \textsc{HM-LDM}($p=1,\delta^2=\text{identifiable}$) & .562	&.502	&.400   &.392 	&.228	&.205   & .527  &  .485\\
\textsc{HM-LDM}($p=1,\delta^2=10^3$) & .439	&.386	& .308 &.303	&.176	& .133 & .405  &  .377
\\
\midrule
\textsc{LDM-Re}-$(\ell^2)^2$ &.546	&.370   &.393   &.211  	&.231    &.137 	 &.497     &.327 	  \\
\textsc{\modelabbrv}($p=2,\delta^2=\text{identifiable}$) &.539 	&.506  &.384 & .373	&.217 &.183	 &.507   &	.504	 
\\
\textsc{HM-LDM}($p=2,\delta^2=10^3$) &.240 	&.133  &.206  & .119	& .116 &.056	 & .232  &.209		 
\\
\bottomrule    
\end{tabular}%
}
\end{table*}

\textbf{Complexity analysis:} The \textsc{HM-LDM} framework requires the computation of the node pairwise distance matrix and consequently scales prohibitively as $\mathcal{O}(N^2)$ in time and space. Fortunately, there are various ways of scaling \textsc{HM-LDM} for the analysis of large-scale networks. One way is through unbiased estimators of the log-likelihood given by Eq. \eqref{eq:prob_adj}. This is possible through random sampling a set of network nodes $S$ (per iteration) and taking a gradient step based on the log-likelihood of the block defined by the sampled node-set, returning an $\mathcal{O}(S^2)$ space and time complexity. Another option is through the case-control approach \cite{case_control} scaling on the number of network edges as $\mathcal{O}(E)$. Lastly, the Hierarchical Block Distance Model (\textsc{HBDM}) \cite{nakis2022hierarchical} is an attractive option where gradient steps over the model parameters are based on a hierarchical approximation of the likelihood of the whole network. The \textsc{HBDM} model scales linearithmicly as $\mathcal{O}(N\log N)$ both in space and time while also offering hierarchical characterizations of structures at multiple scales.

\section{Conclusion and future work}\label{sec:conclusion}
In this paper, we have proposed the \textsc{\modelabbrv} that reconciles network embedding and latent community detection. The approach utilizes both the normal and squared Euclidean distance model where the latter integrated the non-negativity constrained Eigenmodel with the Latent Distance Model. We demonstrated that the model could be constrained to the simplex without losing expressive power. The reduced simplex provides unique representations, ultimately resulting in hard clustering of nodes to communities when the simplex is sufficiently shrunk. Notably, the proposed \textsc{\modelabbrv} combines network homophily and transitivity properties with latent community detection enabling explicit control of soft and hard assignment through the volume of the induced simplex. We observed favorable link prediction performance in regimes in which the \textsc{\modelabbrv} provides unique representations while enabling the ordering of the adjacency matrix in terms of prominent latent communities. Finally, we showed the ability of the model to extract correct community structures across multiple networks and showcased how the analysis extends to bipartite networks. Future work should compare the performance of \textsc{\modelabbrv} against classical non-embedding methods such as the Degree Corrected Stochastic Block Model (\textsc{DC-SBM})\cite{karrer2011stochastic} or the Mixed Membership Stochastic Block Model (\textsc{MM-SBM}) \cite{JMLR:v9:airoldi08a}. Such a comparison is of particular interest since \textsc{DC-SBM} accounts for degree heterogeneity while \textsc{MM-SBM} for soft assignments, two important properties of \textsc{HM-LDM}.

\section*{Acknowledgements}
We would like to thank the reviewers for the constructive feedback and their insightful comments. We would also like to thank Sune Lehmann, Louis Boucherie, Lasse Mohr Mikkelsen, and Giorgio Giannone for the useful and fruitful discussions. We gratefully acknowledge the Independent Research Fund Denmark for supporting this work [grant number: 0136-00315B].

%
%
\bibliographystyle{splncs03.bst}
\bibliography{main}

\begin{thebibliography}{10}
\providecommand{\url}[1]{\texttt{#1}}
\providecommand{\urlprefix}{URL }

\bibitem{JMLR:v9:airoldi08a}
Airoldi, E.M., Blei, D.M., Fienberg, S.E., Xing, E.P.: Mixed membership
  stochastic blockmodels. J Mach Learn Res  9(65),  1981--2014 (2008)

\bibitem{nmf1}
Ball, B., Karrer, B., Newman, M.E.J.: An efficient and principled method for
  detecting communities in networks. CoRR  abs/1104.3590 (2011)

\bibitem{louvainNE-wsdm20}
Bhowmick, A.K., Meneni, K., Danisch, M., Guillaume, J.L., Mitra, B.:
  {LouvainNE}: Hierarchical louvain method for high quality and scalable
  network embedding. In: WSDM. pp. 43--51 (2020)

\bibitem{expon_fam_emb}
{\c{C}}elikkanat, A., Malliaros, F.D.: Exponential family graph embeddings. In:
  {AAAI}. pp. 3357--3364 (2020)

\bibitem{com_metrics}
Chakraborty, T., Dalmia, A., Mukherjee, A., Ganguly, N.: Metrics for community
  analysis: A survey (2016)

\bibitem{node2vec-kdd16}
Grover, A., Leskovec, J.: {Node2Vec}: Scalable feature learning for networks.
  In: KDD. pp. 855--864 (2016)

\bibitem{survey_hamilton_leskovec}
Hamilton, W.L., Ying, R., Leskovec, J.: Representation learning on graphs:
  Methods and applications. {IEEE} Data Eng. Bull.  40(3),  52--74 (2017)

\bibitem{handcock2007model}
Handcock, M.S., Raftery, A.E., Tantrum, J.M.: Model-based clustering for social
  networks. J R Stat Soc Ser A Stat Soc.  170(2),  301--354 (2007)

\bibitem{doi:10.1198/016214504000001015}
Hoff, P.D.: Bilinear mixed-effects models for dyadic data. JASA  100(469),
  286--295 (2005)

\bibitem{hoff2007modeling}
Hoff, P.D.: Modeling homophily and stochastic equivalence in symmetric
  relational data (2007)

\bibitem{exp1}
Hoff, P.D., Raftery, A.E., Handcock, M.S.: Latent space approaches to social
  network analysis. JASA  97(460),  1090--1098 (2002)

\bibitem{nmf5}
Huang, K., Sidiropoulos, N.D., Swami, A.: Non-negative matrix factorization
  revisited: Uniqueness and algorithm for symmetric decomposition. IEEE Trans.
  Signal Process  62(1),  211--224 (2014)

\bibitem{868688}
{Jianbo Shi}, {Malik}, J.: Normalized cuts and image segmentation. IEEE
  Transactions on Pattern Analysis and Machine Intelligence  22(8),  888--905
  (2000)

\bibitem{karrer2011stochastic}
Karrer, B., Newman, M.E.: Stochastic blockmodels and community structure in
  networks. Physical review E  83(1),  016107 (2011)

\bibitem{kingma2017adam}
Kingma, D.P., Ba, J.: Adam: A method for stochastic optimization (2017)

\bibitem{KRIVITSKY2009204}
Krivitsky, P.N., Handcock, M.S., Raftery, A.E., Hoff, P.D.: Representing degree
  distributions, clustering, and homophily in social networks with latent
  cluster random effects models. Social Networks  31(3),  204 -- 213 (2009)

\bibitem{SymmNMF}
Kuang, D., Ding, C., Park, H.: Symmetric nonnegative matrix factorization for
  graph clustering. In: SDM (2012)

\bibitem{lee99}
Lee, D.D., Seung, H.S.: Learning the parts of objects by nonnegative matrix
  factorization. Nature  401,  788--791 (1999)

\bibitem{snapnets}
Leskovec, J., Krevl, A.: {SNAP Datasets}: {Stanford} large network dataset
  collection (Jun 2014)

\bibitem{nmf4}
Mao, X., Sarkar, P., Chakrabarti, D.: On mixed memberships and symmetric
  nonnegative matrix factorizations. In: ICML. vol.~70 (2017)

\bibitem{fb_nets}
Mucha, P., Porter, M.: Social structure of facebook networks. Physica A:
  Statistical Mechanics and its Applications  391,  4165–4180 (08 2012)

\bibitem{nakis2022hierarchical}
Nakis, N., Çelikkanat, A., Jørgensen, S.L., Mørup, M.: A hierarchical block
  distance model for ultra low-dimensional graph representations (2022)

\bibitem{newman}
Newman, M.E.J.: The structure and function of complex networks. SIAM Review
  45(2),  167--256 (2003)

\bibitem{10.5555/2980539.2980649}
Ng, A.Y., Jordan, M.I., Weiss, Y.: On spectral clustering: Analysis and an
  algorithm. In: Proceedings of the 14th International Conference on Neural
  Information Processing Systems: Natural and Synthetic. p. 849–856. NIPS'01,
  MIT Press, Cambridge, MA, USA (2001)

\bibitem{deepwalk-perozzi14}
Perozzi, B., Al-Rfou, R., Skiena, S.: Deepwalk: Online learning of social
  representations. In: KDD. p. 701–710 (2014)

\bibitem{netsmf-www2019}
Qiu, J., Dong, Y., Ma, H., Li, J., Wang, C., Wang, K., Tang, J.: {NetSMF}:
  Large-scale network embedding as sparse matrix factorization. In: WWW (2019)

\bibitem{netmf-wsdm18}
Qiu, J., Dong, Y., Ma, H., Li, J., Wang, K., Tang, J.: Network embedding as
  matrix factorization: Unifying {DeepWalk}, {LINE}, {PTE}, and {Node2Vec}. In:
  WSDM. pp. 459--467 (2018)

\bibitem{case_control}
Raftery, A.E., Niu, X., Hoff, P.D., Yeung, K.Y.: Fast inference for the latent
  space network model using a case-control approximate likelihood. Journal of
  Computational and Graphical Statistics  21(4),  901--919 (2012)

\bibitem{ryan2017bayesian}
Ryan, C., Wyse, J., Friel, N.: Bayesian model selection for the latent position
  cluster model for social networks. Network Science  5(1),  70--91 (2017)

\bibitem{NNSED}
Sun, B.J., Shen, H., Gao, J., Ouyang, W., Cheng, X.: A non-negative symmetric
  encoder-decoder approach for community detection. In: CIKM (2017)

\bibitem{line}
Tang, J., Qu, M., Wang, M., Zhang, M., Yan, J., Mei, Q.: {LINE}: Large-scale
  information network embedding. In: WWW. pp. 1067--1077 (2015)

\bibitem{MNMF}
Wang, X., Cui, P., Wang, J., Pei, J., Zhu, W., Yang, S.: Community preserving
  network embedding. In: AAAI (2017)

\bibitem{nmf2}
Wind, D.K., Mørup, M.: Link prediction in weighted networks. In: 2012 IEEE
  Int. Workshop MLSP. pp. 1--6 (2012)

\bibitem{nmf3}
Yang, J., Leskovec, J.: Overlapping community detection at scale: A nonnegative
  matrix factorization approach. In: WSDM (2013)

\bibitem{gnn_bad}
Yang, L., Gu, J., Wang, C., Cao, X., Zhai, L., Jin, D., Guo, Y.: Toward
  unsupervised graph neural network: Interactive clustering and embedding via
  optimal transport. In: ICDM (2020)

\bibitem{GRL-survey-ieeebigdata20}
{Zhang}, D., {Yin}, J., {Zhu}, X., {Zhang}, C.: Network representation
  learning: A survey. IEEE Trans. Big Data  6(1) (2020)

\bibitem{prone-ijai19}
Zhang, J., Dong, Y., Wang, Y., Tang, J., Ding, M.: Prone: Fast and scalable
  network representation learning. In: IJCAI (2019)

\end{thebibliography}

\end{document}